\documentclass[aps,10pt,pre,floatfix,twocolumn,superscriptaddress]{revtex4-2}

\usepackage{color}
\usepackage{graphicx}
\usepackage{subfigure,amsmath, amsthm, amssymb}
\usepackage[colorlinks,linkcolor=blue,citecolor=blue]{hyperref}

\newcommand{\be}{\begin{equation}}
\newcommand{\ee}{\end{equation}}
\newcommand{\bea}{\begin{eqnarray}}
\newcommand{\eea}{\end{eqnarray}}

\begin{document}

\title{Mpemba effect in driven granular gases: role of distance measures}

\author{Apurba Biswas}
\email{apurbab@imsc.res.in}
\affiliation{The Institute of Mathematical Sciences, C.I.T. Campus, Taramani, Chennai 600113, India}
\affiliation{Homi Bhabha National Institute, Training School Complex, Anushakti Nagar, Mumbai 400094, India}
\author{V . V. Prasad}
\email{prasad.vv@cusat.ac.in}
\affiliation{Department of Physics, Cochin University of Science and Technology,  Kochi 682022, India}
\author{R. Rajesh} 
\email{rrajesh@imsc.res.in}
\affiliation{The Institute of Mathematical Sciences, C.I.T. Campus, Taramani, Chennai 600113, India}
\affiliation{Homi Bhabha National Institute, Training School Complex, Anushakti Nagar, Mumbai 400094, India}

\begin{abstract}
Mpemba effect refers to the counterintuitive effect where a system which is initially further from the final steady state equilibrates faster than an identical system that is initially closer.  The closeness to the final state is defined in terms of a distance measure. For driven granular systems, the Mpemba effect has been illustrated in terms of an ad-hoc measure of mean kinetic energy as the distance function. In this paper,  by studying four different distance measures based on the mean kinetic energies as well as velocity distribution,  we show that the Mpemba effect depends on the definition of the measures.
\end{abstract}

\maketitle
\section{\label{sec:1}Introduction}

Usual experience tells us that a hotter object takes a longer time to cool than a similar object initially at a lower temperature. Such phenomena are well described using Newtons's cooling law where rate of cooling is directly proportional to the difference in temperature between the object and its surrounding. Mpemba effect refers to a counterintuitive phenomenon where a system which is initially at a higher temperature equilibrates faster than a similar system which is initially at a lower intermediate temperature, when both the systems are quenched to the same low  temperature.  This effect was first observed in the case of water where freezing time was found to be lower for water that is initially at higher temperature~\cite{Mpemba_1969,wojciechowski1988freezing,Mirabedin-evporation-2017, vynnycky-convection:2015,katz2009hot,david-super-cooling-1995}. 
However, the effect is not limited to water and has been experimentally shown to exist  in many other physical systems such as  magnetic alloys~\cite{chaddah2010overtaking}, clathrate hydrates~\cite{paper:hydrates}, polylactides~\cite{Polylactide}, carbon nanotube resonators~\cite{greaney2011-cnt-mpemba} and colloidal systems~\cite{kumar2020exponentially,kumar2021anomalous,bechhoefer2021fresh}, thus suggesting that the Mpemba effect is a much more general anomalous relaxation phenomenon that can be studied in numerous physical systems. Theoretical studies on the Mpemba effect have focused on  spin systems~\cite{PhysRevLett.124.060602,Klich-2019,klich2018solution,das2021should,PhysRevE.104.044114,teza2021relaxation}, systems undergoing phase transitions~\cite{das2021should,holtzman2022landau,zhang2022theoretical}, Markovian systems with only a few states~\cite{Lu-raz:2017,PhysRevResearch.3.043160}, systems of single particle diffusing in a potential~\cite{Walker_2021,Busiello_2021,lapolla2020faster,degunther2022anomalous}, active systems~\cite{schwarzendahl2021anomalous}, spin glasses~\cite{SpinGlassMpemba}, molecular gases in contact with a thermal reservoir~\cite{moleculargas,gonzalez2020mpemba,PhysRevE.104.064127,PhysRevE.105.054140}, quantum systems~\cite{PhysRevLett.127.060401} and granular systems~\cite{Lasanta-mpemba-1-2017,Torrente-rough-2019,mompo2020memory,biswas2020mpemba,biswas2021mpemba,biswas2022mpemba,megias2022mpemba}. 

The  protocol followed to test for the existence of the Mpemba effect considers two identical systems which are prepared in different initial states. Both the systems are then quenched to a  common final state, and the Mpemba effect is said to be present if the system that is initially further away from the final state equilibrates faster. 
To quantify which initial state is further away as well as crossing of trajectories, a distance measure  for points in the phase space has to be defined.  

For systems relaxing to thermal equilibrium,  the distance to the final equilibrium state  is measured in terms of the probabilities of the different states~\cite{Lu-raz:2017}. It was argued that any distance measure satisfying the following properties should result in a unique definition of the Mpemba effect: (1) As the system relaxes toward thermal equilibrium, the  distance function should decrease with time, (2) for three temperatures $T_h>T_c>T_b$, the distance from $T_b$ is larger for $T_h$ compared  to $T_c$, i.e., the distance measure should be a monotonically increasing function of temperature and (3) the distance function should be a continuous, convex function of probability distribution~\cite{Lu-raz:2017}. Let $\pi$ denote the equilibrium Boltzmann distribution, and $p_i(t)$ the probability of state $i$ at time $t$. Then, examples of such distances are ``entropic distance" $D_e=\sum_i (p_i(t)-\pi_i)E_i/T_b +p_i(t) \ln p_i(t)- \pi_i \ln \pi_i$~\cite{Lu-raz:2017}, total variation distance $L_1(t)=\sum_i |p_i(t)-\pi_i|$~\cite{kumar2020exponentially} and Kullback-Leibler (KL) divergence defined as $D_{KL}(t)=\sum_i p_i(t) \ln (p_i(t)/\pi_i)$~\cite{kullback1951information}.


The calculation of the above definitions of distance rely on knowing the probability distribution $p(t)$ at all times.  Analytical calculation of $p(t)$ is possible only for exactly solvable problems which, in the context of the Mpemba effect, are restricted to simple single particle systems~\cite{kumar2020exponentially,kumar2021anomalous,bechhoefer2021fresh} and to systems with only a few states~\cite{Lu-raz:2017}. For interacting many-particle systems and more so in the context of out of equilibrium systems, the distance measures defined in the probability space are, however, inaccessible through direct measurements in experiments or computationally expensive to measure in simulations. The more natural choices for the distance measures that have been used in experiments are moments of the distribution which are directly observable like mean energy per spin~\cite{SpinGlassMpemba,das2021should}, magnetization~\cite{chaddah2010overtaking,das2021should}, granular temperature~\cite{Lasanta-mpemba-1-2017,Torrente-rough-2019,mompo2020memory,biswas2020mpemba,biswas2021mpemba,biswas2022mpemba,megias2022mpemba}, etc. In addition, the generalization to far from equilibrium systems, like driven granular systems is not clear. There are no conclusive studies about the correspondence or similarity between the directly observable measures and the distance measures defined in probability space.  Thus, it is not evident whether the Mpemba effect that has been established using these proxy measures for distances is unique or its existence depends on the distance measure used.


To address the above issue, we study the dependence of the Mpemba effect on the choice of different measures in driven granular systems, a  prototypical interacting, many-particle far from equilibrium  system. The advantage of granular system is that it allows exact analysis for two-point correlation functions, at least in the linearized regime. Being an athermal system, granular temperature or the mean kinetic energy of the system has been used to track the evolution of the system with higher granular temperature being considered further away from the final steady state~\cite{Lasanta-mpemba-1-2017,mompo2020memory,Torrente-rough-2019,biswas2020mpemba,biswas2021mpemba,biswas2022mpemba}. 
However, it is not known a priori whether mean kinetic energy correctly predicts the distance of the initial states from the final steady states.  In the following, we summarize the several results that demonstrates the existence of the Mpemba effect in granular systems. For a system of smooth monodispersed particles~\cite{Lasanta-mpemba-1-2017,mompo2020memory,biswas2020mpemba}, the Mpemba effect is due to the coupling of the translational granular temperature with the excess kurtosis of the velocity distribution function or correlations between the velocities of the different particles. In the case of rough granular gas~\cite{Torrente-rough-2019}, the Mpemba effect is due to the coupling of granular temperatures defined for translational and rotational degrees of freedom. In these cases, for the Mpemba effect to exist, however the initial states need to be different from steady states. Existence of the Mpemba effect for evolution of systems from initial steady states  could be realised  in the context of binary inelastic gases or when inelastic gases are driven anisotropically. For the case of binary gas, Mpemba effect was traced to the energy-exchange between the energies of the smaller and bigger particles~\cite{biswas2020mpemba}, while 
in anisotropically driven gas~\cite{biswas2021mpemba,biswas2022mpemba}, the Mpemba effect is due to the coupling between the granular temperatures along $x$ and $y$ directions. However, in all the above mentioned analysis, the time evolution of the system is projected onto only one of the variables which is the total granular temperature defined by the second moment of velocity distribution of the system.

In this paper, in addition to total energy, we introduce other measures such as Manhattan measure ($\mathcal{L}_1$), Euclidean measure ($\mathcal{L}_2$) and KL divergence which can describe the evolution of the system in the  phase space of all the relevant variables. We perform the analysis in the setup of anisotropically driven granular gas as well as driven inelastic Maxwell gas~\cite{biswas2021mpemba,biswas2022mpemba}. We derive the criteria for the existence of the Mpemba effect with the various measures and determine the region of phase space which shows the Mpemba effect. We show that these phase diagrams are non-universal in the sense that they depend on the measure used.

The remainder of the paper is organized as follows. We first define the model of anisotropically driven granular and Maxwell gases in Sec.~\ref{sec:2}. In Sec.~\ref{sec:3}, we derive the time evolution of  the two-point correlation functions, exactly for the Maxwell gas, and in the linearized regime for the granular gas. It allows us to exactly calculate their  steady state values which characterize the steady state. In Sec.~\ref{sec:4}, we define the Mpemba effect and the various measures that we use to characterize the distance between steady states. In Sec.~\ref{sec:5}, we present the results for the Mpemba effect using the various measures. We compare the various measures for the Mpemba effect in terms of their phase diagrams. Section~\ref{sec:6} contains the conclusion and its implications.

\section{\label{sec:2}Models}

In this paper, we analyze two models for the driven granular gas in two dimensions: inelastic Maxwell model and the hard disc granular gas model. Consider a collection of identical,  inelastic particles. The velocities of these particles change in time due to inelastic binary collisions that are momentum conserving. The new velocities $\boldsymbol{v}'_i$ and $\boldsymbol{v}'_j$ of the two colliding particles $i$ (velocity $\boldsymbol{v}_i$) and $j$ (velocity $\boldsymbol{v}_j$) after the collision is given by
\bea
\begin{split}
\boldsymbol{v}'_i=\boldsymbol{v}_i -\alpha[(\boldsymbol{v}_i-\boldsymbol{v}_j).\boldsymbol{\hat{e}}]\boldsymbol{\hat{e}},  \\
\boldsymbol{v}'_j=\boldsymbol{v}_j +\alpha[(\boldsymbol{v}_i-\boldsymbol{v}_j).\boldsymbol{\hat{e}}]\boldsymbol{\hat{e}},
\end{split} \label{eq:collision}
\eea
where
\bea
\alpha=\frac{1+r}{2},
\eea
and $r$ is the coefficient of restitution and $\boldsymbol{\hat{e}}$ is the unit vector along the line joining the centers of the particles at contact. In addition, the particles are driven anisotropically at a constant rate, $\lambda_d$. We describe the driving for each model separately.

{\bf  Inelastic Maxwell gas:}
In the inelastic Maxwell gas model, the collision rates of the particles are independent of the relative velocities of the colliding particles. Thus, the spatial  degrees of freedom are  irrelevant. These simplifications lead to the model being analytically tractable, yet keeping essential characteristics of more complicated realistic models.

In addition to the collision rules described by Eq.~(\ref{eq:collision}), each particle is driven with rate $\lambda_d$. When particle $i$ is driven, its velocity $\boldsymbol{v}_i$  is changed to $\boldsymbol{v}'_i$, given by
\bea
v'_{ix,y}=-r_{wx,y}v_{ix,y}+\eta_{ix,y}, \label{eqn driving}
\eea
where $r_{wx,y}$ are scalar parameters associated with the driving along the $x$ and $y$ directions, and $\boldsymbol{\eta}$ is a noise taken from a fixed distribution $\phi({\boldsymbol{\eta}})$ with a finite second moment denoted by
\bea
\sigma^2_{k}=\int^{\infty}_{-\infty}d\eta_k \eta^2_k \phi(\boldsymbol{\eta}), \quad k=x,y.
\eea
When $\sigma^2_x\neq \sigma^2_y$, the driving is anisotropic and leads to particles having an anisotropic velocity distribution.
The physical motivations for the form of driving in Eq.~(\ref{eqn driving}) may be found in Refs.~\cite{Prasad:18,Prasad:19}.

Let $P(\boldsymbol{v},t)$ denote the probability that a randomly chosen particle has velocity $\boldsymbol{v}$ at time $t$. Its time evolution is given by
\begin{widetext}
\begin{align}
\frac{d P(\boldsymbol{v},t)}{dt}&=\lambda_c \int \int \int d\boldsymbol{\hat{e}} d\boldsymbol{v}_1 d\boldsymbol{v}_2 P(\boldsymbol{v}_1,t) P(\boldsymbol{v}_2,t) \delta(\boldsymbol{v}_1- \alpha[(\boldsymbol{v}_1-\boldsymbol{v}_2).\boldsymbol{\hat{e}}]\boldsymbol{\hat{e}}- \boldsymbol{v})-\lambda_c  P(\boldsymbol{v},t)
 \nonumber \\
& +\lambda_d \int \int d\boldsymbol{\eta} d\boldsymbol{v}_1 \Phi(\boldsymbol{\eta}) P(\boldsymbol{v}_1,t)\delta[-r_{wx}v_{1x}+ \eta_x-v_x ] \delta[-r_{wy}v_{1y}+ \eta_y-v_y] 
-\lambda_d P(\boldsymbol{v},t), \label{eq:time ev} 
\end{align}
\end{widetext}
where the first and second terms on the right hand side describe the gain and loss terms due to collisions while the third and fourth terms describe the gain and loss terms due to driving.

{\bf  Hard disc granular gas:}
In the hard disc granular gas,  the collision rates are proportional to the relative velocities of the colliding particles, as expected for ballistic motion. The time evolution of the velocity distribution function $f(\boldsymbol{v},t)$, defined as the number density of particles having velocity $\boldsymbol{v}$ at time $t$ is described using the Enskog-Boltzmann equation~\cite{Noije:98}
\begin{align}
\frac{\partial}{\partial t}f(\boldsymbol{{v}},t)=\chi I(f,f) + \Big(\frac{\sigma^2_{x}}{2} \frac{\partial^2}{\partial v^2_x} + \frac{\sigma^2_{y}}{2} \frac{\partial^2}{\partial v^2_y}\Big) f(\boldsymbol{{v}},t), \label{boltzmann eq}
\end{align}
where $\chi$ is the pair correlation function~\cite{brilliantov2010kinetic}, $I(f,f)$ (see Appendix~\ref{appendix 2}, Eq.~(\ref{collision integral}) for details) is the collision integral which accounts for the gain and loss terms in Eq.~(\ref{eq:time ev}) due to collisions, and $\sigma^2_{x}$ and $\sigma^2_{y}$ are the variances or strengths of the white noise along the $x$- and $y$-directions respectively. 

The driving terms in Eq.~(\ref{boltzmann eq}), corresponding to the diffusion-like terms, are related to the driving for the Maxwell gas given in Eq.~(\ref{eqn driving}). By Taylor expanding the terms related to driving in Eq.~(\ref{eq:time ev}), and truncating for small noise, it is easily shown that the driving terms in Eq.~(\ref{boltzmann eq}) corresponds to the special case $r_{wx}=r_{wy}=1$~\cite{Prasad:18,Prasad:19,biswas2021mpemba}. 

Note that we have considered a spatially homogeneous system to describe the system such that  spatial degrees of freedom are ignored. Moreover, we have also assumed molecular chaos hypothesis to use product measure for the joint velocity distribution function in the collision integral.

\section{\label{sec:3}Characterizing the steady states}
In this section, we define the relevant two point correlation functions for both the models. The  time evolutions of these two point correlations were already derived in Refs.~\cite{biswas2021mpemba,biswas2022mpemba}. However,  we summarize the derivations  and compute their steady state values as they form an integral part of the present analysis. Sections  \ref{sec:3a} and \ref{sec:3b} contain the derivations for the inelastic Maxwell model and the hard disc granular gas model respectively. The text follows closely the text in Refs.~\cite{biswas2021mpemba,biswas2022mpemba}.
\subsection{\label{sec:3a}Inelastic Maxwell model}
For the  inelastic Maxwell model, we are interested in the  following two-point correlation functions:
\bea
E_x(t)&=&\frac{1}{N}\sum^{N}_{i=1} \langle v^2_{ix}(t) \rangle,  \nonumber \\
E_y(t)&=&\frac{1}{N}\sum^{N}_{i=1} \langle v^2_{iy}(t) \rangle, \nonumber \\
E_{xy}(t)&=&\frac{1}{N }\sum^{N}_{i=1} \langle v_{ix}(t) v_{iy}(t) \rangle, \nonumber \\
C_{x}(t)&=&\frac{1}{N (N-1)}\sum^{N}_{i=1} \sum_{\substack{j=1 \\ j \neq i}}^{N} \langle v_{ix}(t) v_{jx}(t) \rangle,  \label{2 point correlations} \\
C_{y}(t)&=&\frac{1}{N (N-1)}\sum^{N}_{i=1} \sum_{\substack{j=1 \\ j \neq i}}^{N} \langle v_{iy}(t) v_{jy}(t) \rangle, \nonumber \\
C_{xy}(t)&=&\frac{1}{N (N-1)}\sum^{N}_{i=1} \sum_{\substack{j=1 \\ j \neq i}}^{N} \langle v_{ix}(t) v_{jy}(t) \rangle,  \nonumber 
\eea
where $E_x(t)$ and $E_y(t)$ denote the mean kinetic energies of the particles along $x$ and $y$ directions respectively. $E_{xy}(t)$ denote the correlations between $v_x$ and $v_y$ of the same particle whereas $C_x(t)$, $C_y(t)$ and $C_{xy}(t)$ denote the velocity-velocity correlations between pairs of particles. The time evolution of these correlation functions is derived using Eq.~(\ref{eq:time ev}) and is compactly written in the form
\begin{align}
\frac{d\boldsymbol{\tilde{\Sigma}}(t)}{dt} =\boldsymbol{\tilde{R}}  \boldsymbol{\tilde{\Sigma}}(t) + \boldsymbol{\tilde{D}},
\label{two-point-correlation-matrix-evolution}
\end{align}
where the column vectors $\boldsymbol{\tilde{\Sigma}}(t)$ and $\boldsymbol{\tilde{D}}$ are given by:
\begin{align}
\begin{split}
&\boldsymbol{\tilde{\Sigma}}(t)=[E_x(t),E_y(t),E_{xy}(t),C_x(t),C_y(t),C_{xy}(t)]^T,  \\
&\boldsymbol{\tilde{D}}=[\lambda_d \sigma^2_x,\lambda_d \sigma^2_y, 0, 0, 0, 0]^T.  \label{matrices sigma and D}
\end{split}
\end{align}

The components of the matrix $\boldsymbol{\tilde{R}}$ are given in Appendix~\ref{appendix 1}. In the steady state, in the thermodynamic limit, the velocity-velocity correlations and the correlation between $v_x$ and $v_y$ of the same particle, i.e., $E_{xy}$ vanish as shown in Ref.~\cite{biswas2022mpemba}. Moreover, if these correlations are zero for the initial state, then it remains zero for all times. In that case, we only write for the time evolution of the non-zero mean kinetic energies, i.e., $E_x$ and $E_y$ in a compact form as
\begin{equation}
\frac{d\boldsymbol{\Sigma}(t)}{dt}=~\boldsymbol{R}\boldsymbol{\Sigma}(t)+\boldsymbol{S},\label{time-ev}
\end{equation}
where
\begin{align}
&\boldsymbol{\Sigma}(t)=\begin{bmatrix}
E_{x}(t), 
E_{y}(t)
\end{bmatrix}^T, \\
&\boldsymbol{S}=\begin{bmatrix}
\lambda_d \sigma^2_x, 
\lambda_d \sigma^2_y
\end{bmatrix}^T,
\end{align}
and  $\boldsymbol{R}$ is a  $2\times2$ matrix, whose entries are given by
\begin{align}
\begin{split}
R_{11}=&\frac{3}{4}\lambda_c \alpha^2 -\lambda_c \alpha -\lambda_d(1-r^2_{wx}) , \quad R_{12}=\frac{\lambda_c}{4}\alpha^2,\\ 
R_{22}=&\frac{3}{4}\lambda_c \alpha^2 -\lambda_c \alpha -\lambda_d(1-r^2_{wy}), \quad R_{21}=\frac{\lambda_c}{4}\alpha^2 . \\ 
\end{split}
\end{align}

We will be using a different set of variables than $E_x$ and $E_y$ for the analysis. We define the new set of variables, namely the total kinetic energy, $E_{tot}$, and the difference of energies, $E_{dif}$, as:
\begin{eqnarray}
E_{tot}=E_x+E_y,
\label{E1 define}\\
E_{dif}=E_x-E_y.
\label{E2 define}
\end{eqnarray}
The time evolution equations for $E_{tot}$ and $E_{dif}$ can be expressed, starting from Eq.~(\ref{time-ev}), as
\begin{equation}
\frac{d\boldsymbol{E}(t)}{dt}=-~\boldsymbol{\chi}\boldsymbol{E}(t)+\boldsymbol{D}, \label{time ev E}
\end{equation}
where
\begin{align}
&\boldsymbol{E}(t)=\begin{bmatrix}
E_{tot}(t), 
E_{dif}(t)
\end{bmatrix}^T, \\
&\boldsymbol{D}=\begin{bmatrix}
\lambda_d (\sigma^2_x+\sigma^2_y), 
\lambda_d (\sigma^2_x-\sigma^2_y)
\end{bmatrix}^T,
\end{align}
and $\boldsymbol{\chi}$ is a $2\times2$ matrix with the components of the matrix given by 
\begin{align}
\begin{split}
\chi_{11}&=\frac{2\lambda_c\alpha(1-\alpha)+\lambda_d(2-r^2_{wx}-r^2_{wy})}{2}, \\
\chi_{22}&=\frac{\lambda_c\alpha(2-\alpha)+\lambda_d(2-r^2_{wx}-r^2_{wy})}{2}, \\
 \chi_{12}&=\frac{\lambda_d(r^2_{wy}-r^2_{wx})}{2},\quad \chi_{21}=\frac{\lambda_d(r^2_{wy}-r^2_{wx})}{2}. \label{appendix : chi}
\end{split} 
\end{align}
Equation~(\ref{time ev E}) can be solved exactly by linear decomposition using the eigenvalues $\lambda_\pm$ of $\boldsymbol{\chi}$:
\bea
\lambda_\pm=&\frac{1}{4}\Big[2 \lambda_d (2-r^2_{wx}-r^2_{wy}) + \alpha \lambda_c(4 -3 \alpha) \nonumber \\
&\pm \sqrt{4 \lambda^2_d(r^2_{wy}-r^2_{wx})^2+ \alpha^4 \lambda^2_c}\Big].
\eea

It is straightforward to show that $\lambda_{\pm}>0$ with $\lambda_+>\lambda_-$. The solution for $E_{tot}(t)$ and $E_{dif}(t)$ is
\begin{align}
\begin{split}
E_{tot}(t)- E^{st}_{tot}&=K_+e^{-\lambda_+t}+ K_-e^{-\lambda_-t}, \\
E_{dif}(t)- E^{st}_{dif}&=L_+e^{-\lambda_+t}+ L_-e^{-\lambda_-t},
\end{split} \label{sol E1 one driven}
\end{align}
where $E^{st}_{tot}$ and $E^{st}_{dif}$ are steady state values of $E_{tot}(t)$ and $E_{dif}(t)$ respectively. The coefficients $K_+, K_-, L_+$ and $L_-$ along with $E^{st}_{tot}$ and $E^{st}_{dif}$  are given in Appendix~\ref{appendix 1}.

\subsection{\label{sec:3b}Hard disc granular gas model}
For the model of driven hard disc granular gas whose evolution is described using Eq.~(\ref{boltzmann eq}), the mean kinetic energies along the $x$ and $y$ directions  are defined as
\bea
E_i(t)=\frac{2}{n}\int d \boldsymbol{v} \frac{1}{2} m v^2_i  f(\boldsymbol{{v}},t),\quad i=x,y,
\eea
where $n=\int d \boldsymbol{v} f(\boldsymbol{{v}},t)$ is the number density, $m$ is the mass of the particles. In order to derive the time evolution of the mean kinetic energies using Eq.~(\ref{boltzmann eq}), we assume an anisotropic form of the Gaussian for the velocity distribution function (which is valid when the system is weakly  inelastic) as described in Ref.~\cite{biswas2021mpemba},
\begin{equation}
f(\boldsymbol{v},t)=\frac{mn}{2\pi \sqrt{E_x(t) E_y(t)}}\exp\left[ -\frac{m v^2_{x}}{2 E_x(t)}-\frac{m v^2_{y}}{2 E_y(t)} \right]. \label{gaussian distribution}
\end{equation}
With this approximation for $f(\boldsymbol{v},t)$, the time evolution of $E_{tot}$ and $E_{dif}$ [defined as in Eqs.~(\ref{E1 define}) and (\ref{E2 define})] form a coupled set of non-linear differential equations and is given by
\bea
\begin{split}
\frac{\partial }{\partial t} E_{tot}(t)=\mathcal{F}(E_{tot},E_{dif}),\\
\frac{\partial }{\partial t}E_{dif}(t)=\mathcal{G}(E_{tot},E_{dif}).
\end{split} \label{time ev}
\eea
The functional forms for $\mathcal{F}(E_{tot},E_{dif})$ and $\mathcal{G}(E_{tot},E_{dif})$ are given in  Appendix~\ref{appendix 2}. In order to perform an exact analysis, we linearize the non-linear equations by considering only the initial states that are close to the final steady state. We define $\delta E_{tot}(t)=E_{tot}(t)- E^{st}_{tot}$ and $\delta E_{dif}(t)=E_{dif}(t) - E^{st}_{dif}$ as the time-dependent deviation of the energies from the stationary state values. The linearized form for the non-linear differential equations [see Eq.~(\ref{time ev})] about the stationary state values obey
\begin{equation}
\frac{d}{dt}\begin{bmatrix}
\delta E_{tot}(t) \\
\delta E_{dif}(t)
\end{bmatrix}=-~\boldsymbol{\chi}\begin{bmatrix}
\delta E_{tot}(t) \\
\delta E_{dif}(t)
\end{bmatrix}, \label{time ev delta T}
\end{equation}
where $\boldsymbol{\chi}$ is a $2\times2$ matrix whose coefficients are given in the Appendix~\ref{appendix 2}. The solutions for $\delta E_{tot}(t)$ and $\delta E_{dif}(t)$ are then given by
\bea
\begin{split}
\delta E_{tot}(t)=K_+ e^{-\lambda_+ t} + K_- e^{-\lambda_- t}, \\
\delta E_{dif}(t)=L_+ e^{-\lambda_+ t} + L_- e^{-\lambda_- t}, \label{time ev delta Tt}
\end{split}
\eea 
where  the coefficients $K_+, K_-, L_+$ and $L_-$ are given by
\bea
\begin{split}
K_+&=\frac{1}{\gamma}\Big[\chi_{12} \delta E_{dif}(0)  - (\lambda_- - \chi_{11}) \delta E_{tot}(0) \Big], \\
K_-&=\frac{1}{\gamma}\Big[-\chi_{12} \delta E_{dif}(0)  + (\lambda_+ - \chi_{11}) \delta E_{tot}(0) \Big], \\
L_+&=\frac{1}{\gamma}\Big[(\lambda_+ - \chi_{11}) \delta E_{dif}(0)  \\
&  - \frac{(\lambda_+ - \chi_{11})(\lambda_- - \chi_{11})}{\chi_{12}} \delta E_{tot}(0) \Big], \\
L_-&=\frac{1}{\gamma}\Big[-(\lambda_- - \chi_{11}) \delta E_{dif}(0)   \\ 
&  + \frac{(\lambda_+ - \chi_{11})(\lambda_- - \chi_{11})}{\chi_{12}} \delta E_{tot}(0) \Big].
\end{split} \label{coefficients}
\eea
Here, $\lambda_\pm$ are the eigenvalues of $\boldsymbol{\chi}$ and $\gamma=\lambda_+-\lambda_-$.

\section{\label{sec:4} Mpemba effect and distance measures in phase space}

We first define the protocol that we will follow for illustrating the Mpemba effect.  Consider two systems $P$ and $Q$ which have identical  parameters except for the driving strengths. Both the systems, in their respective steady states, are then quenched to a common steady state. This is achieved by instantaneously changing the driving strengths of $P$ and $Q$ to the common driving strength of the final steady state, keeping all the other parameters of both the systems fixed. The two initial steady states $P$ and $Q$  differ  in their initial distance (to be appropriately defined) from the final steady state. The evolution of the systems $P$ and $Q$ correspond to two different trajectories in phase space. Then the Mpemba effect is said to exist if the trajectory that was initially at a larger distance from the final steady state approaches the final state faster than the trajectory that was initially at a shorter distance. 
For granular systems, several distance measures have been used to explore the Mpemba effect. We derive the criterion for the existence of the  Mpemba effect for the different measures and for each measure, we illustrate the Mpemba effect. We also ask how much the Mpemba effect depends on the distance measure being used. 

The steady state of the two models for driven granular systems considered in this paper is completely specified by the velocity distribution $P({\boldsymbol{ v}})$. Distance between two probability distributions may be defined in terms of 
an information theoretic quantity known as Kullback-Leibler (KL) divergence~\cite{marconi2013h,PhysRevE.95.052121,megias2020kullback}. 
However, the velocity distribution cannot be solved exactly and therefore the KL divergence becomes unwieldy for studying Mpemba effect, though we will study this numerically. Instead, the steady state of the system has been parametrised by the moments of the velocity. Note that the equations for the two point correlation functions close among themselves for both the Maxwell gas as well as the linearized granular gas. This motivates using the second moments of velocity distribution to characterize the steady states. For the system of anisotropically driven granular gas, the mean kinetic energies, $E_x$ and $E_y$ along $x$- and $y$- directions respectively, are different. Thus, the total energy, $E_{tot}=E_x+E_y$ and the difference of energies, $E_{dif}=E_x-E_y$ serve as the appropriate variable to describe a state of the system. Thus, a steady state is characterized by  $(E_{tot}, E_{dif})$.

In this section, we introduce different measures that have been used to define the distance between two steady states of the system.  When the steady state is defined through $(E_{tot}, E_{dif})$, the distance measures can be defined in terms of the difference of the  total energy of the two states (Sec.~\ref{total energy}), in terms of two dimensional Euclidean distance (Sec.~\ref{euclidean distance}) or two-dimensional Manhattan distance (Sec.~\ref{manhattan distance}) of the phase space variables of the two states. In terms of the velocity distribution, the convergence to the steady state can be characterized through KL divergence (Sec.~\ref{divergence}). The different measures track the temporal evolution of the system as it evolves from an initial state to a final state, and we derive the conditions for the Mpemba effect to exist.   


\subsection{\label{total energy}Total energy}

The most common variable that is used in literature of driven granular gases to describe its state is the mean kinetic energy.  The existence of the Mpemba effect in driven granular systems has been shown in previous studies~\cite{biswas2020mpemba,biswas2021mpemba,biswas2022mpemba} in terms of these variables. In this section, we briefly discuss the condition for the existence of the Mpemba effect in an anisotropically driven granular gas. We consider two systems $P$ and $Q$ whose initial steady states are denoted by $[E^P_{tot}(0),E^P_{dif}(0)]$ and $[E^Q_{tot}(0),E^Q_{dif}(0)]$ respectively. Here, the distance of the initial states compared to the final state is measured in terms of the total energy. The initial steady states for the two systems are prepared such that  $E^P_{tot}(0)>E^Q_{tot}(0)$.  Both the systems are then quenched to a common steady state having the total energy lower than the initial total energies of both the systems.

The Mpemba effect  is present if  the two trajectories $E^P_{tot}(t)$ and $E^Q_{tot}(t)$ cross each other at some finite time $t=\tau$ at which
\begin{equation}
E^P_{tot}(\tau)=E^Q_{tot}(\tau).
\end{equation}

To obtain the value of $\tau$, we equate the total energies of $P$ and $Q$ using either Eq.~(\ref{sol E1 one driven}) or Eq.~(\ref{time ev delta Tt}) depending on the inelastic Maxwell gas or hard disc granular gas respectively to obtain
\begin{equation}
K^P_+e^{-\lambda_+\tau}+ K^P_-e^{-\lambda_-\tau}=K^Q_+e^{-\lambda_+\tau}+ K^Q_-e^{-\lambda_-\tau}, \label{evolution eq at crossing}
\end{equation}
whose solution is
\begin{equation}
\tau=\frac{1}{\lambda_+-\lambda_-}\ln \Big[\frac{K^P_+-K^Q_+}{K^Q_--K^P_-} \Big]. 
\end{equation}

In terms of the parameters of the initial steady states, $\tau$ reduces to
\begin{equation}
\tau=\frac{1}{\lambda_+-\lambda_-} \ln \Big[\frac{\chi_{12}\Delta E_{dif}-(\lambda_--\chi_{11})\Delta E_{tot}}{\chi_{12}\Delta E_{dif}-(\lambda_+-\chi_{11})\Delta E_{tot}}  \Big], \label{crossing time one driven}
\end{equation}
where
\begin{align}
\begin{split}
&\Delta E_{tot}=E^P_{tot}(0)-E^Q_{tot}(0), \\
&\Delta E_{dif}=E^P_{dif}(0)-E^Q_{dif}(0).
\end{split} \label{eq: delta E}
\end{align}
For the Mpemba effect to be present, we require that $\tau >0$. Since $\lambda_+>\lambda_-$, the argument of logarithm in Eq.~(\ref{crossing time one driven}) should be greater than one. 

\begin{figure}
\centering
\includegraphics[width= \columnwidth]{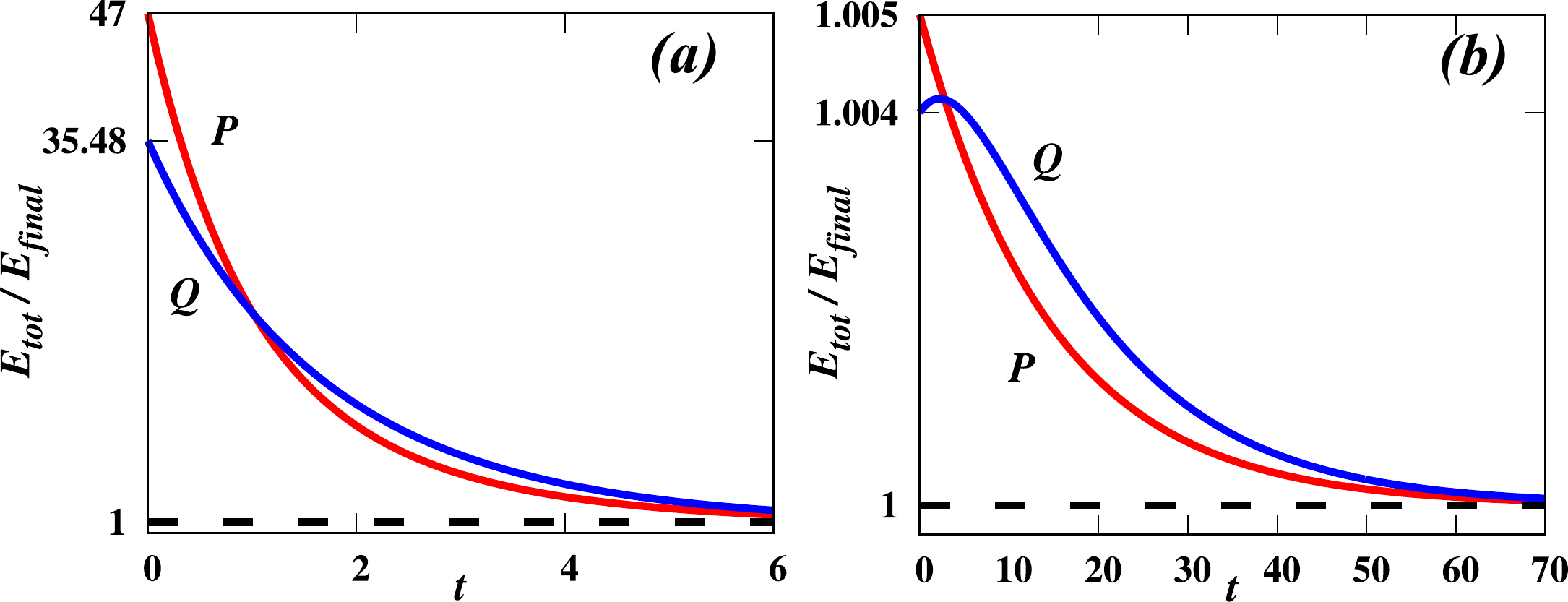} 
\caption{\label{fig1 energy evolution}The time evolution of anisotropically driven (a) inelastic Maxwell and (b) hard disc granular gas is illustrated in terms of the mean kinetic energy, $E_{tot}$ for two systems $P$ and $Q$. The initial conditions for the inelastic Maxwell gas in (a) are $E^P_{tot}(0)=1.148$, $E^Q_{tot}(0)=0.92$, $E^P_{dif}(0)=-0.595$ and $E^Q_{dif}(0)=0.455$. The choice of the other parameters defining the system are $r=0.3$, $r_{wx}=0.88$, $r_{wy}=0.39$, $\sigma_x=0.1$ and $\sigma_y=0.05$. The initial conditions for the hard disc granular gas in (b) are $E^P_{tot}(0)=10.05$, $E^Q_{tot}(0)=10.04$, $E^P_{dif}(0)=2.797$ and $E^Q_{dif}(0)=1.979$. The choice of the other parameters defining the system are $r=0.65$, $m=1$, $n=0.02$, $E^{st}_{tot}=10.0$ and $E^{st}_{dif}=2.82$. $P$ relaxes to the steady state faster than $Q$, though it is initially at a larger distance compared to the final steady state. }
\end{figure}

Figure~\ref{fig1 energy evolution} illustrates the existence of the Mpemba effect where the trajectories of the initial states leading to final steady state are defined in terms of the total energy and their crossing time is given by Eq.~(\ref{crossing time one driven}).

\subsection{\label{euclidean distance}Euclidean Distance}

Mean kinetic energy as a measure of distance has the possible issue that the trajectories may appear to cross in this one-dimensional projection, though in the two dimensional phase they do not cross. More natural definitions are to use Euclidean or Manhattan distances.  As the initial and the final states are defined by the pair of variables $\{E_{tot}(t), E_{dif}(t)\}$, we can define an Euclidean measure for the trajectory connecting the two states as
\bea
\mathcal{L}_2(t)=\sqrt{(E_{tot}(t)-E^{st}_{tot})^2 + (E_{dif}(t)-E^{st}_{dif})^2}.
\eea

Here, the initially ``hotter" system or equivalently the system which is initially farther from the final steady state has an initially larger $\mathcal{L}_2$ compared to the ``colder'' system. For this measure, we define the Mpemba effect as follows. Let us consider two systems $P$ and $Q$ such that $P$ is initially at a larger distance from the final steady state compared to $Q$, i.e., $\mathcal{L}^P_2(0)>\mathcal{L}^Q_2(0)$. Here, the systems $P$ and $Q$ are identical in all respect except for the pair of driving strengths $(\sigma_x,\sigma_y)$ that is required to prepare the systems in their initial steady states. Then both the systems are quenched to a common steady state by applying a same pair of driving strengths. In this case, the Mpemba effect is said to exist if  the two trajectories for the systems $P$ and $Q$ quantified in terms of $\mathcal{L}^P_2(t)$ and $\mathcal{L}^Q_{2}(t)$ cross each other at some finite time $t=\tau$ at which
\begin{equation}
\mathcal{L}^P_{2}(\tau)=\mathcal{L}^Q_{2}(\tau).
\end{equation}

To obtain the value of $\tau$, we equate the Euclidean measures for the systems $P$ and $Q$ at $t=\tau$ as
\bea
&\sqrt{(E^P_{tot}(\tau)-E^{st}_{tot})^2 + (E^P_{dif}(\tau)-E^{st}_{dif})^2} \nonumber \\
&=\sqrt{(E^Q_{tot}(\tau)-E^{st}_{tot})^2 + (E^Q_{dif}(\tau)-E^{st}_{dif})^2}.
\eea

For the inelastic Maxwell gas,  using Eq.~(\ref{sol E1 one driven})  we obtain the crossing times as
\bea
\tau^{\pm}=\frac{1}{(\lambda_+-\lambda_-)} \ln\left[\frac{2a}{d\pm \sqrt{d^2-4ab}} \right], \label{maxwell crossing time}
\eea
where,
\bea
\begin{split}
a&=\big[(K^P_+)^2+(L^P_+)^2-(K^Q_+)^2-(L^Q_+)^2\big], \\
b&=\big[(K^P_-)^2+(L^P_-)^2-(K^Q_-)^2-(L^Q_-)^2\big],\\
d&=2\big[K^Q_+ K^Q_- + L^Q_+ L^Q_- - K^P_+ K^P_- - L^P_+ L^P_-\big].
\end{split} \label{constants a,b and d}
\eea
A similar expression for the crossing time [see Eq.~(\ref{maxwell crossing time})] is also obtained for the case of hard disc granular gas for which the constants $\lambda_{\pm}$, $a$, $b$, and $d$  are computed using Eq.~(\ref{coefficients}).

\begin{figure}
\centering
\includegraphics[width= \columnwidth]{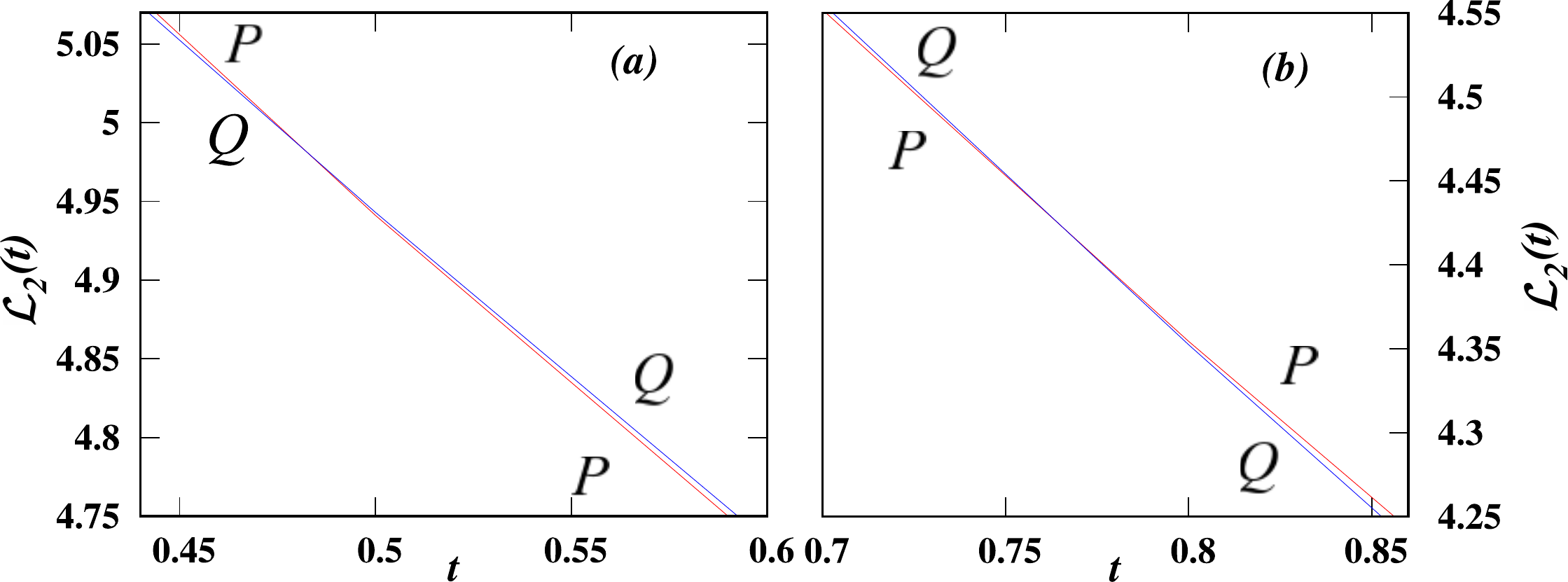} 
\caption{\label{fig2 multiple crossing}The time evolution of the  anisotropically driven inelastic Maxwell gas in terms of the Euclidean measure, $\mathcal{L}_2(t)$ for two systems $P$ and $Q$ with initial conditions  $\mathcal{L}^P_2(0)=6.34$ and $\mathcal{L}^Q_2(0)=6.15$, shows two crossings as illustrated in (a) and (b) for the different times. The multiple crossing times are obtained using Eq.~(\ref{maxwell crossing time}). The choice of the other parameters defining the system are $r=0.1$, $r_{wx}=0.95$, $r_{wy}=0.39$, $\sigma_x=1.6$ and $\sigma_y=1.1$. }
\end{figure}

\begin{figure}
\centering
\includegraphics[width= \columnwidth]{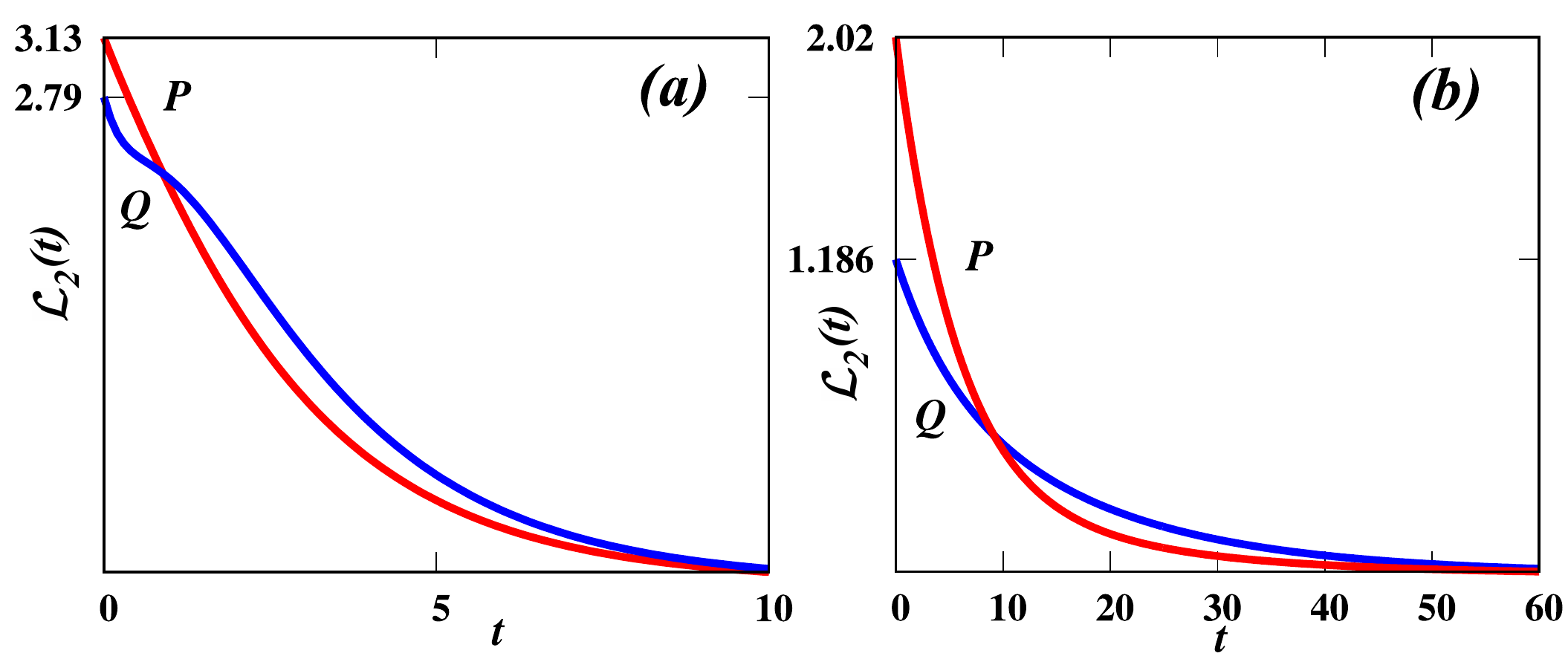} 
\caption{\label{fig3 evolution euclidean}The time evolution of anisotropically driven (a) inelastic Maxwell and (b) hard disc granular gas is illustrated in terms of Euclidean measure, $\mathcal{L}_2(t)$ for two systems $P$ and $Q$. The initial conditions for the inelastic Maxwell gas in (a)  are $\mathcal{L}^P_2(0)=3.13$ and $\mathcal{L}^Q_2(0)=2.79$. The choice of the other parameters defining the system are $r=0.3$, $r_{wx}=0.95$, $r_{wy}=0.39$, $\sigma_x=1.6$ and $\sigma_y=1.1$.  The initial conditions for the hard disc granular gas in (b)  are $\mathcal{L}^P_2(0)=2.02$ and $\mathcal{L}^Q_2(0)=1.186$. The choice of the other parameters defining the system are $r=0.3$, $m=1$ and $n=0.02$. $P$ relaxes to the steady state faster than $Q$, though its initial Euclidean distance from the final steady state is larger.}
\end{figure}

Note that there are two possibilities for the crossing time in Eq.~(\ref{maxwell crossing time}). Figure~\ref{fig2 multiple crossing} illustrates such a scenario where both the crossings are present. However, the presence of two crossings will eventually lead to no Mpemba effect. Thus, we are interested in only those cases or the initial conditions where there exists only one crossing of the trajectories of the initial states leading to the final steady state.
For the Mpemba effect to be present, we require that $\tau^+ >0$ or $\tau^{-}>0$, but not both positive. Since $\lambda_+>\lambda_-$, the argument of logarithm in Eq.~(\ref{maxwell crossing time}) should be greater than one for $+ (-)$ and less than one for $- (+)$. Figure~\ref{fig3 evolution euclidean} illustrates the existence of the Mpemba effect where the trajectories of the initial states leading to final steady state are defined in terms of the Euclidean measure and their crossing time is given by Eq.~(\ref{maxwell crossing time}).

\subsection{\label{manhattan distance}Manhattan Distance}

Similar to the Euclidean measure, we can define another measure for the time evolution of the trajectory connecting the initial and the final states in terms of Manhattan distance. In the plane of $\{E_{tot}(t), E_{dif}(t)\}$, the Manhattan measure is defined as
\bea
\mathcal{L}_1(t)=|E_{tot}(t)-E^{st}_{tot}| + |E_{dif}(t)-E^{st}_{dif}|. 
\eea

Here, the system having larger $\mathcal{L}_1(0)$ at time $t=0$ is termed the ``hotter" system while compared to the initially ``colder" system having a comparatively smaller $\mathcal{L}_1(0)$. For this measure, we define the Mpemba effect as follows. Let us consider two systems $P$ and $Q$ such that $P$ is initially at a larger distance from the final steady state compared to $Q$ in terms of Manhattan measure, i.e., $\mathcal{L}^P_1(0)>\mathcal{L}^Q_1(0)$. Here, the systems $P$ and $Q$ are identical except for the pair of driving strengths ($\sigma_x$, $\sigma_y$). Both the systems are then subjected to same pair of driving strengths. As a result, both the systems are driven   to a common steady state. Then the Mpemba effect is said to exist if  the two trajectories for the systems $P$ and $Q$ quantified in terms of $\mathcal{L}^P_1(t)$ and $\mathcal{L}^Q_1(t)$ cross each other at some finite time $t=\tau$ at which
\begin{equation}
\mathcal{L}^P_1(\tau)=\mathcal{L}^Q_1(\tau),
\end{equation}
or equivalently,
%
\bea
&|(E^P_{tot}(\tau)-E^{st}_{tot})| + |(E^P_{dif}(\tau)-E^{st}_{dif})| \nonumber \\
&=|(E^Q_{tot}(\tau)-E^{st}_{tot})| + |(E^Q_{dif}(\tau)-E^{st}_{dif})|. \label{equate manhattan distance}
\eea

Now, using Eq.~(\ref{sol E1 one driven}) for the case of inelastic Maxwell gas and solving Eq.~(\ref{equate manhattan distance}), we obtain eight different solutions for the crossing time as
\bea
\begin{split}
\tau_{c,1}&=\frac{1}{\gamma}\ln\Big(\frac{-K^P_++K^Q_++L^P_++L^Q_+}{K^P_--K^Q_--L^P_--L^Q_-}\Big), \\ \label{crossing time 1}
\tau_{c,2}&=\frac{1}{\gamma}\ln\Big(\frac{-K^P_+-K^Q_+-L^P_++L^Q_+}{K^P_-+K^Q_-+L^P_--L^Q_-}\Big), \\
\tau_{c,3}&=\frac{1}{\gamma}\ln\Big(\frac{-K^P_+-K^Q_++L^P_+-L^Q_+}{K^P_-+K^Q_--L^P_-+L^Q_-}\Big),\\
\tau_{c,4}&=\frac{1}{\gamma}\ln\Big(\frac{-K^P_++K^Q_+-L^P_+-L^Q_+}{K^P_--K^Q_-+L^P_-+L^Q_-}\Big),\\
\tau_{c,5}&=\frac{1}{\gamma}\ln\Big(\frac{-K^P_+-K^Q_++L^P_++L^Q_+}{K^P_-+K^Q_--L^P_--L^Q_-}\Big),\\
\tau_{c,6}&=\frac{1}{\gamma}\ln\Big(\frac{-K^P_++K^Q_+-L^P_++L^Q_+}{K^P_--K^Q_-+L^P_--L^Q_-}\Big),\\
\tau_{c,7}&=\frac{1}{\gamma}\ln\Big(\frac{-K^P_++K^Q_++L^P_+-L^Q_+}{K^P_--K^Q_--L^P_-+L^Q_-}\Big),\\
\tau_{c,8}&=\frac{1}{\gamma}\ln\Big(\frac{-(K^P_++K^Q_++L^P_++L^Q_+)}{K^P_-+K^Q_-+L^P_-+L^Q_-}\Big). \label{crossing time manhattan}
\end{split} 
\eea

\begin{figure}
\centering
\includegraphics[width= \columnwidth]{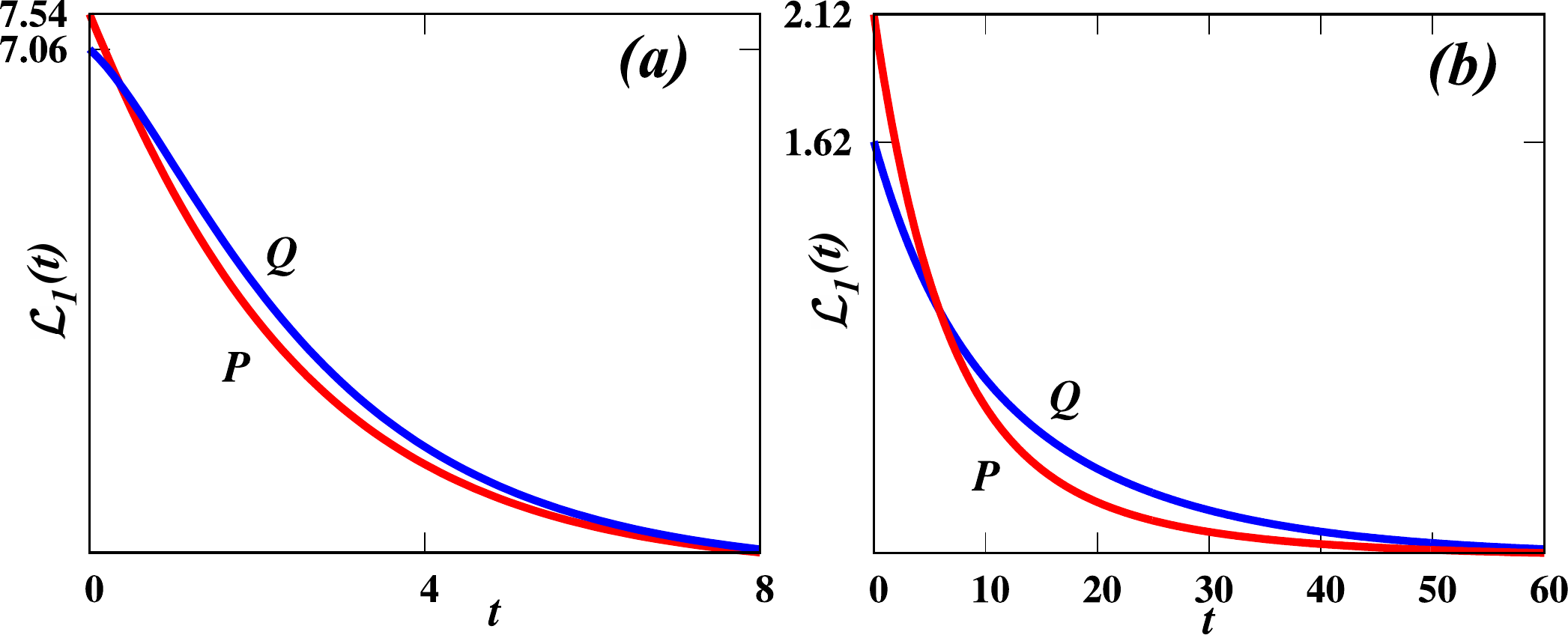} 
\caption{\label{fig4 evolution manhattan}The time evolution of anisotropically driven (a) inelastic Maxwell and (b) hard disc granular gas is illustrated in terms of Manhattan measure, $\mathcal{L}_1(t)$ for two systems $P$ and $Q$. The initial conditions for the inelastic Maxwell gas in (a)  are $\mathcal{L}^P_1(0)=7.54$ and $\mathcal{L}^Q_1(0)=7.06$. The choice of the other parameters defining the system are $r=0.9$, $r_{wx}=0.88$, $r_{wy}=0.3$, $\sigma_x=1.6$ and $\sigma_y=1.0$.  The initial conditions for the hard disc granular gas in (b)  are $\mathcal{L}^P_1(0)=2.12$ and $\mathcal{L}^Q_1(0)=1.62$. The choice of the other parameters defining the system are $r=0.3$, $m=1$ and $n=0.02$. $P$ relaxes to the steady state faster than $Q$, though its initial Manhattan distance from the final steady state is larger.}
\end{figure}

For the Mpemba effect to be present, there should be odd number of 
crossings. Thus, we look for those initial conditions in the phase space that lead to odd number of solutions to the Eq.~(\ref{equate manhattan distance}).  Figure~\ref{fig4 evolution manhattan} illustrates the existence of the Mpemba effect where the trajectories of the initial states leading to final steady state are defined in terms of the Manhattan measure. 


\subsection{\label{divergence}KL Divergence}
In this section, we define the Mpemba effect in terms of an information theoretic quantity known as Kullback-Leibler (KL) divergence. The measure is defined as 
\bea
D_{KL}(t)=\int d\boldsymbol{v} P(\boldsymbol{v},t) \ln \Big(\frac{P(\boldsymbol{v},t)}{P^{st}(\boldsymbol{v})}\Big), \label{KL divergence}
\eea
where $P(\boldsymbol{v},t)$ is the instantaneous velocity distribution of the particles at any time $t$ and $P^{st}(\boldsymbol{v})$ is the final steady state velocity distribution function. The above quantity is not a true measure of geometric distance as it is not symmetric between two given distribution functions. But it is a good candidate for the study of relaxation dynamics of an arbitrary initial state to a given reference steady state for the following reasons: (a) it is a monotonically non-increasing function of time as has been shown in earlier studies~\cite{marconi2013h,PhysRevE.95.052121,megias2020kullback}, (b) it provides information regarding the measure of deviation and its temporal evolution of any two initial states from the final reference state.

Although being a good candidate for the study of relaxation dynamics,  the above measure is not much fruitful in the context of granular gases. It is because Eq.~(\ref{KL divergence}) requires to have information regarding the time evolution of the velocity distribution function at any time $t$ which is not analytically feasible in the case of granular gases. The previous theoretical studies on velocity distribution of driven granular gases attempts to find the velocity distribution of the steady state for different contexts~\cite{PhysRevE.95.022115,PhysRevE.95.032909,Prasad:18,Prasad:19,biswas2020asymptotic}. But in all the cases, it was not possible to derive the form of instantaneous velocity distribution at an instant of time $t$ even for the simplest model of inelastic Maxwell gas.

As a result, we use numerical methods to compute the KL divergence. We briefly discuss the procedure for the numerical computation of KL divergence for the case of inelastic Maxwell gas. Starting from a random configuration of velocities for $N$ particles, the system is evolved to an initial steady state corresponding to an initial pair of driving strengths $(\sigma_x, \sigma_y)$. The pair of driving strengths acting on the particles is then changed to those of the desired final steady state. The system evolves in time as follows. At each time step, either two particles collide or a particle is driven, based on the corresponding rates. At each time step we measure the time evolution of the velocity distribution, $P(\boldsymbol{v},t)$ of the $N$ particles. The distributions are averaged over $10^4$ realizations. The final steady state velocity distribution is measured separately and is averaged over $10^5$ different realizations. Thus, having information about $P(\boldsymbol{v},t)$ and $P^{st}(\boldsymbol{v})$, we can then numerically determine the time evolution of KL divergence using Eq.~(\ref{KL divergence}).


We now discuss the notion of a ``hotter" and a ``colder" system using the measure of KL divergence when two systems are quenched to a final steady state. The system having larger divergence at $t=0$ is referred to as the ``hotter" system as it is farther off from the final steady state whereas the system having comparatively smaller divergence is referred to as the ``colder" system. Similar to the previous definitions of the Mpemba effect for the other measures, if the hotter system equilibrates faster than the colder system to the final steady state, the Mpemba effect is said to exist. 
\begin{figure}
\includegraphics[width= \columnwidth]{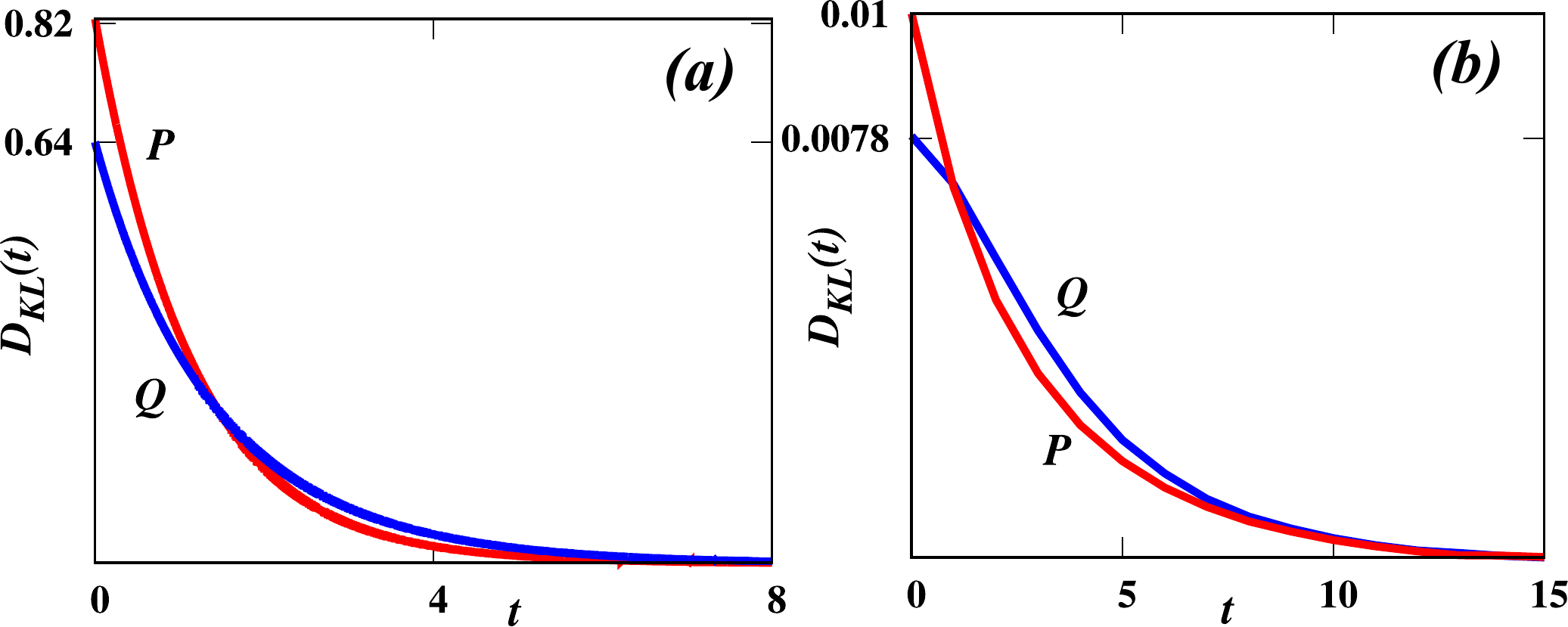} 
\caption{\label{fig5 evolution kl}The time evolution of anisotropically driven (a) inelastic Maxwell and (b) hard disc granular gas is illustrated in terms of KL divergence measure, $D_{KL}(t)$ for two systems $P$ and $Q$. The initial conditions for the inelastic Maxwell gas in (a)  are $D^P_{KL}(0)=0.82$ and $D^Q_{KL}(0)=0.64$. The choice of the other parameters defining the system are $r=0.3$, $r_{wx}=0.88$, $r_{wy}=0.39$, $\sigma_x=1.5$ and $\sigma_y=0.9$.  The initial conditions for the hard disc granular gas in (b)  are $D^P_{KL}(0)=0.01$ and $D^Q_{KL}(0)=0.0078$. The choice of the other parameters defining the system are $r=0.65$, $m=1$, $n=0.02$, $\sigma_x=6.62$ and $\sigma_y=0.026$. $P$ relaxes to the steady state faster than $Q$, though its initial KL divergence with respect to the final steady state is larger. }
\end{figure}

Figure~\ref{fig5 evolution kl} illustrates the existence of the Mpemba effect where the trajectories of the initial states leading to final steady state are defined in terms of the KL divergence.

 \section{\label{sec:5}Comparison between various measures}
 
We now compare the effect of different distance measures on the Mpemba effect. For a given initial condition in terms of intrinsic system parameters and the pair of driving strengths, we look for the existence of the Mpemba effect using the different distance measures.

\begin{figure}
\centering
\includegraphics[width= \columnwidth]{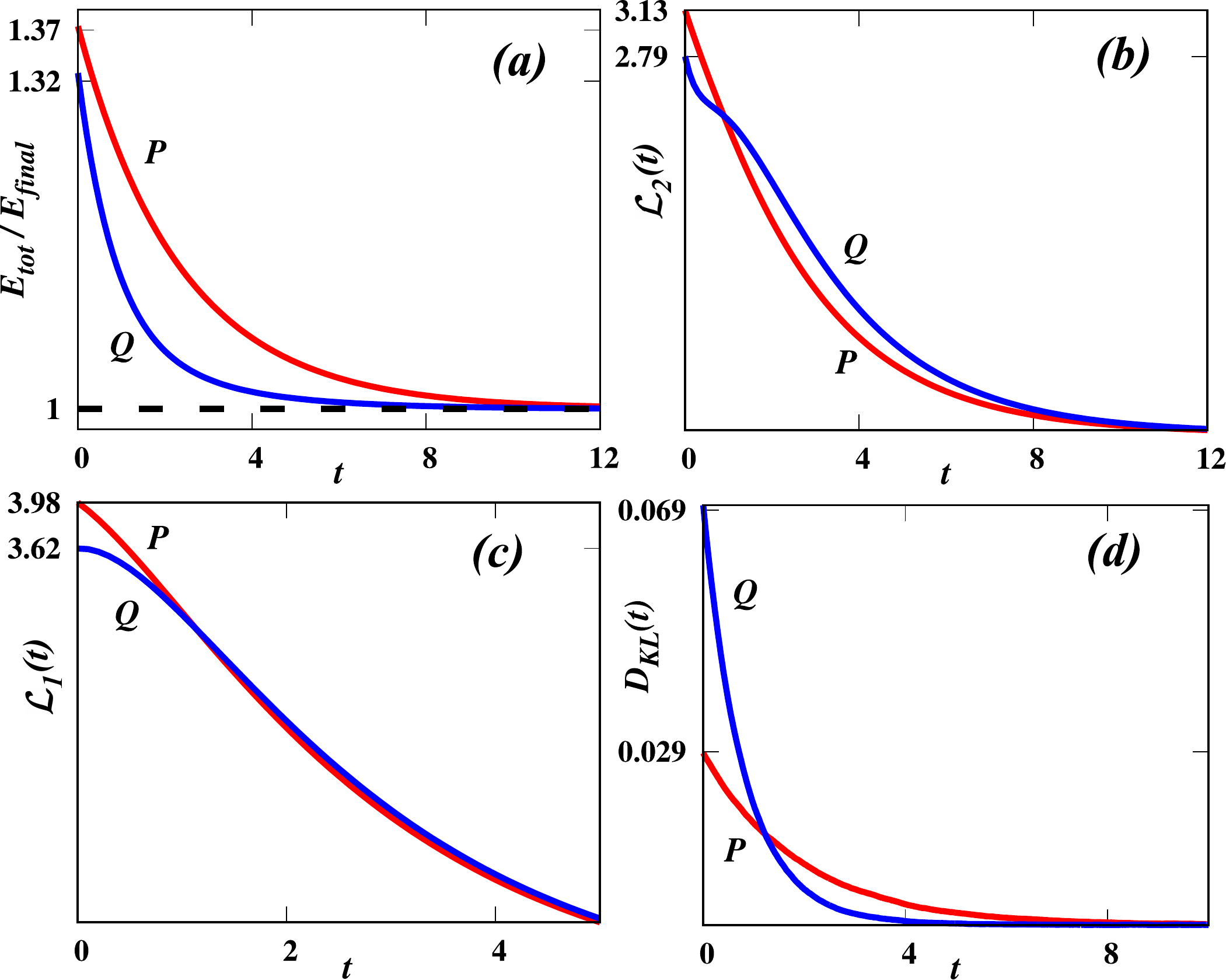} 
\caption{\label{contrasting evolutions}The time evolution of the two systems $P$ and $Q$ for an  anisotropically driven inelastic Maxwell gas  is illustrated for the following measures: (a) total energy, (b) Euclidean measure, (c) Manhattan measure and for (d) KL divergence. The initial conditions for the various measures are identical and are given in terms of the driving strengths for systems $P$ and $Q$ as $(\sigma^P_x=1.9, \sigma^P_y=1.2)$ and $(\sigma^Q_x=1.55, \sigma^Q_y=2.0)$. The choice of the other parameters defining the system are $r=0.3$, $r_{wx}=0.95$, $r_{wy}=0.39$, $\sigma_x=1.6$ and $\sigma_y=1.1$. The existence of the Mpemba effect and the notion of ``hot" and ``cold" system in terms of distance from the final steady state for a given pair of initial conditions is not unique among the various measures.}
\end{figure}

We first show that for the same initial condition, different measures can give different results, as illustrated in Fig.~\ref{contrasting evolutions}  for the inelastic Maxwell gas. In the example shown, the Mpemba effect is absent for distance measure corresponding to the total energy distance [see Fig.~\ref{contrasting evolutions}(a)] which it is present for the other three measures [see Fig.~\ref{contrasting evolutions}(b)--(d)]. In addition to the non-uniqueness of the Mpemba effect, even the notion of ``hot" and ``cold" system in terms of distance from the final steady state for a pair of initial conditions is not unique among the various definitions.  What is initially hotter (shown in red) in Fig.~\ref{contrasting evolutions}(a)--(c) is initially colder (shown in blue) in Fig.~\ref{contrasting evolutions}(d)

Given that there is non-uniqueness among the definitions, we now check whether all definitions show the Mpemba effect as well as whether there is any overlap in the phase space regions that corresponds to the Mpemba effect for the different measures.
The phase space is labeled by the free parameters: final steady states in terms of ($E^{st}_{tot},~E^{st}_{dif}$), initial steady states of $P$ and $Q$, i.e., ($E^{P}_{tot},~E^{P}_{dif}$) and ($E^{Q}_{tot},~E^{Q}_{dif}$) for a fixed value of other intrinsic parameters such as $r$, $r_{wx}$ and $r_{wy}$. However, we notice that  the equations for the correlation functions [see Eqs.~(\ref{sol E1 one driven}) and (\ref{time ev delta T})] are linear equations in the differences in energies $\delta E_{tot}=E_{tot}-E^{st}_{tot}$, $\delta E_{dif}=E_{dif}-E^{st}_{dif}$  for both $P$ and $Q$, giving four variables. We also note that the equations for the correlation function, and hence the crossing times, are invariant if all the $\delta E$s are scaled by the same factor, thus reducing the number of factors by one. To make it a two dimensional phase diagram, we fix the initial values of $\delta E^P_{tot}$ and $\delta E^P_{dif}$  for $P$, and determine the phase diagram in terms of ($\delta E^Q_{dif}/\delta E^P_{tot}$) and ($\delta E^Q_{tot}/\delta E^P_{tot})$.
\begin{figure}
\includegraphics[width= \columnwidth]{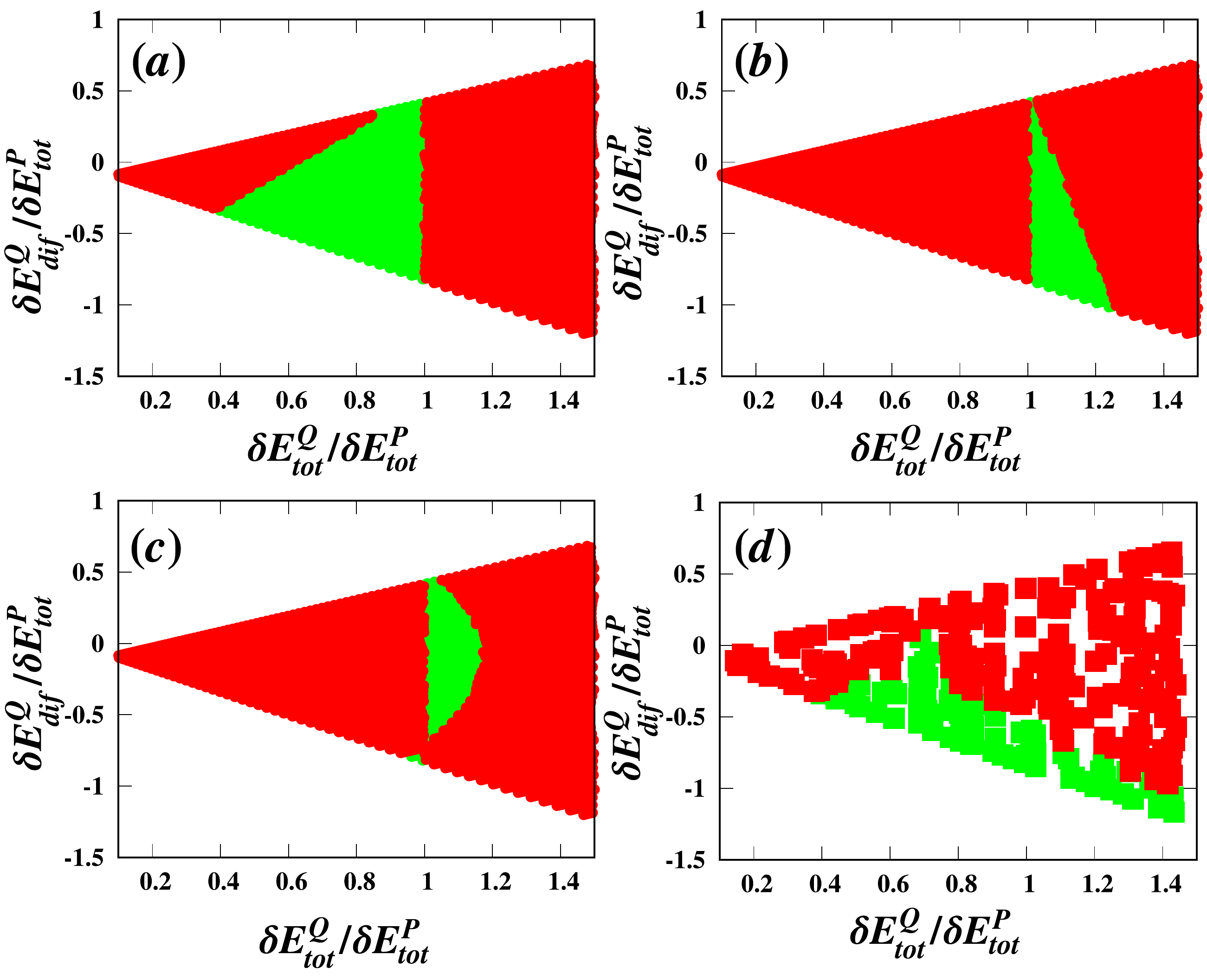} 
\caption{\label{phase diagram maxwell}The phase diagram in the ($\delta E^Q_{dif}/\delta E^P_{tot}$)-($\delta E^Q_{tot}/\delta E^P_{tot})$ plane shows the existence of the Mpemba effect in the driven inelastic Maxwell gas for the use of different distance measures: (a) total energy, (b) Manhattan, (c) Euclidean, and (d) KL-divergence. The red (green) region corresponds to absence (presence) of Mpemba effect, while the white regions are not accessible.  The choice of the parameters defining the system are $r=0.4$, $r_{wx}=0.44$, $r_{wy}=0.95$, $\delta E^P_{tot}=1.00$ and $\delta E^P_{dif}/\delta E^P_{tot}=0.53$, which are kept constant across all the various measures. }
\end{figure}
\begin{figure}
\includegraphics[width= \columnwidth]{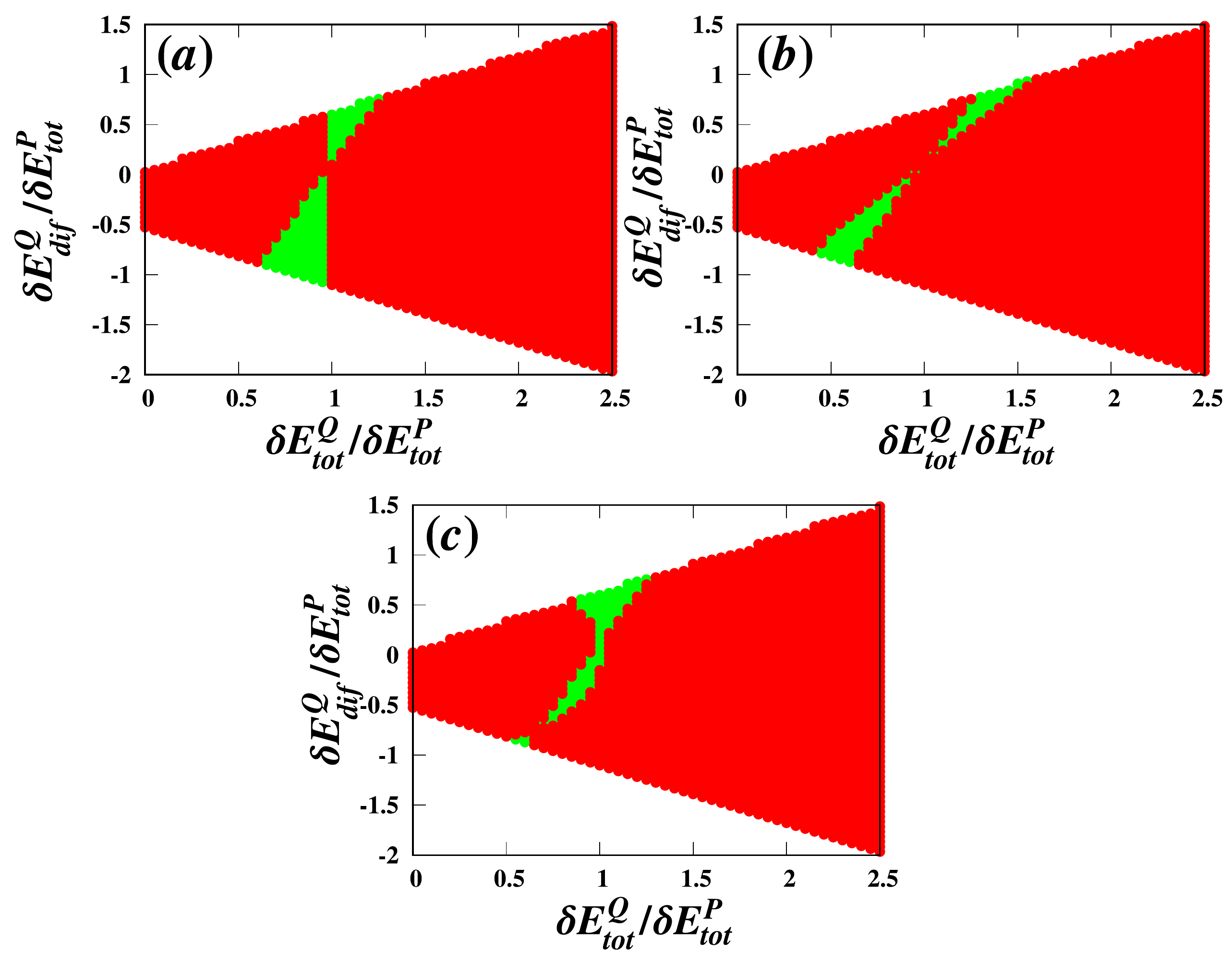} 
\caption{\label{phase diagram real gas} The phase diagram in the ($\delta E^Q_{dif}/\delta E^P_{tot}$)-($\delta E^Q_{tot}/\delta E^P_{tot})$ plane shows the existence of the Mpemba effect in the driven hard disc granular gas for the use of different distance measures: (a) total energy, (b) Manhattan, and (c) Euclidean. The red (green) region corresponds to absence (presence) of Mpemba effect, while the white regions are not accessible.  The choice of the parameters defining the system are $n=0.02$, $\sigma=1$, $m=1$, $r=0.1$, $\delta E^{P}_{tot}=1.00$ and $\delta E^{P}_{dif}/\delta E^P_{tot}=0.11$, which are kept constant across all the various measures.  }
\end{figure}

The phase diagrams for the measures: total energy, Manhattan and Euclidean distance, are obtained using the exact criteria derived for the existence of the Mpemba effect, both for the inelastic Maxwell gas and hard disc granular gas (linearized analysis). However, for the case of KL divergence, the phase diagram is obtained using discrete set of points sampled from the phase space and checked individually for the existence of the Mpemba effect through the analysis of evolution of the trajectories  of systems $P$ and $Q$. The KL divergence is used only for the Maxwell gas as in the case of granular gas, it is not possible to make a direct comparison with the linearized theories. Figures~\ref{phase diagram maxwell} and \ref{phase diagram real gas} show the phase diagram of the Mpemba effect in terms of various measures for the case of inelastic Maxwell gas and hard disc granular gas respectively. The colored region corresponds to the allowed parameter space, while the green (red) regions correspond to the Mpemba effect being present (absent).  It is clear that all the different distance measures show the Mpemba effect. However, the phase boundary of the initial conditions that show the Mpemba effect is very different for the various measures. We conclude that the usual measures used for driven granular gases lead to a non-universal definition of the Mpemba effect.

 \section{\label{sec:6}Summary and conclusion}
 
To summarize, we studied the Mpemba effect in the driven inelastic Maxwell  and the driven granular gases using different definitions of distances between points in phase space. These definitions were motivated by the choices that have been used earlier in the literature for establishing the presence of the Mpemba effect in granular systems. We studied the Mpemba effect in terms of the following measures: total kinetic energy (as has been used earlier),  Euclidean measure ($\mathcal{L}_2$), Manhattan measure ($\mathcal{L}_1$) and KL divergence. While the first three are based on average kinetic energy in different directions, KL divergence is based on the probability distribution of velocity. The analysis was performed for the anisotropically driven  gases~\cite{biswas2021mpemba,biswas2022mpemba}. 


We derived the criteria for the existence of the Mpemba effect with the various measures and showed the existence of the Mpemba effect for all the different choices of distance measures. However, the phase diagrams for the different measures are not the same with the presence or absence of the Mpemba effect for a particular phase space point depending on the measure, resulting in a non-unique definition of the Mpemba effect.  Moreover, the notion of `hot' and `cold' initial states is different for the different choices of distance measures from the final steady state. 

In a Markov process, the relaxation to the steady state at large times is dominated by the first ``excited" state of the transition matrix, i.e., $P(C, t)= P_1(C) + a_2 \exp(-t/\tau) P_2(C)+ \ldots$, where $P(C,t)$ is the probability distribution at time $t$, $P_1$ is the steady state and $P_2$ is the first excited state. The quantities $\tau$, $P_1$ and $P_2$ depend only on the transition matrix, and independent of the initial states, while the pre-factor $a_2$ is determined by the overlap of the initial steady state with the first excited state. The system with smaller value of $|a_2|$ will equilibrate faster~\cite{Lu-raz:2017}.  The same is true for the time evolution of an arbitrary observable $f(C, t)$ as it also follows a similar structure as $P(C,t)$. Therefore, if system $P$ approaches equilibrium faster according to KL-divergence, then we will find similar relaxation behaviour  with total energy as the distance measure. However, what is cold and what is hot may not be the same with both measures (see Fig.~\ref{contrasting evolutions}(a) and \ref{contrasting evolutions}(d) for an example), thus giving contradictory results for the Mpemba effect.  Since it is difficult to come up with an observable $f(C,t)$ that has the same characteristics as KL-divergence, we generalised the total variation distance  $\left|P(C, t)-P_1(C)\right|$ to the distances $\mathcal{L}_1$ and $\mathcal{L}_2$ used in this paper, based on the mean energies.

The ambiguity of characterising the Mpemba effect in granular systems is in contrast to the unique identification of the Mpemba effect for the case of single particle Markov systems, irrespective of the choice of distance measure. This is due to the existence of a common criteria across all the measures which is given in terms of the coefficient $a_2$~\cite{Lu-raz:2017} associated with the second eigenfunction in the eigenspectrum analysis of probability distribution function. However,  it is very  challenging to calculate $a_2$ for an interacting many particle system like the granular system considered in this paper. The natural choices are observables like kinetic energy, rather than probability distributions,  which are easier to track both in experiments and calculations. However, these choices for distance measures may not necessarily decrease monotonically with time, a requirement that was put forward in Ref.~\cite{Lu-raz:2017}. Based on the results in this paper, the characterization of the Mpemba effect with such ad-hoc distance measures should be done with caution.

 \appendix

\section{ Time evolution of two point correlations for inelastic Maxwell gas \label{appendix 1}}
In this appendix, we solve for the time evolution of the various two-point correlations for the inelastic Maxwell gas. The two-point correlations that we are interested in are discussed in Eq.~(\ref{2 point correlations}) whose time evolution is given by  $\frac{d\boldsymbol{\tilde{\Sigma}}(t)}{dt} =\boldsymbol{\tilde{R}}  \boldsymbol{\tilde{\Sigma}}(t) + \boldsymbol{\tilde{D}}$ where the matrices $\boldsymbol{\tilde{\Sigma}}$ and $\boldsymbol{\tilde{D}}$ are given in Eq.~(\ref{matrices sigma and D}). The form of the matrix $\boldsymbol{\tilde{R}}$ is given by
\begin{widetext}
\begin{flalign}
\tilde{\boldsymbol{R}}\!= 
&\!\left[\begin{array}{cccccc} 
\frac{A_{1}(N-1)}{N}+A^{xx}_4 &\frac{A_2(N-1)}{N} &0 & \frac{-A_1(N-1)}{N} &\frac{ -A_2(N-1)}{N} & 0  \\
\frac{A_2(N-1)}{N} & \frac{A_{1}(N-1)}{N}+A^{yy}_4 &0 & \frac{-A_2(N-1)}{N} & \frac{-A_1(N-1)}{N} &0 \\
0 & 0 & \frac{-A_3(N-1)}{N}+A^{xy}_4 &0 &0 &\frac{A_3(N-1)}{N} \\
-\frac{A_1}{N} & -\frac{A_2}{N} & 0 & \frac{A_1}{N}+2A^x_5& \frac{A_2}{N} &0 \\
-\frac{A_2}{N} & -\frac{A_1}{N} & 0 &\frac{A_2}{N} & \frac{A_1}{N}+2A^y_5 & 0 \\
0 & 0 & \frac{A_3}{N} & 0 & 0 & -\frac{A_3}{N} +A^x_5+A^y_5
\end{array}\right],
\label{eq:Rmatrix for any N}
\end{flalign}
\end{widetext}
where the constants $A_1, A_2, A_3, A^{ij}_4, A^{i}_5$, where $i, j\in (x,y)$ are given by 
 \bea
 \begin{split}
 &A_1=\frac{3}{4}\lambda_c \alpha^2-\lambda_c \alpha,\quad  A_2=\frac{\lambda_c \alpha^2}{4}, \\
 &A_3=\lambda_c \alpha(1-\frac{\alpha}{2}),\quad A^{ij}_4=-\lambda_d(1-r_{wi}r_{wj}),\\
 &A^i_5=-\lambda_d(1+r_{wi}), \quad \textnormal{where}~ i,j \in (x,y). \label{matrix constants}
 \end{split}
 \eea
In the thermodynamic limit $N\rightarrow \infty$, the matrix $\tilde{\boldsymbol{R}}$ simplifies to
\begin{flalign}
&\boldsymbol{\tilde{R}}\!=\nonumber \\  
&\!\left[\begin{array}{cccccc} 
A_{1}+A^{xx}_4 &A_2 &0 & -A_1 &-A_2 & 0  \\
A_2 & A_{1}+A^{yy}_4 &0 & -A_2 & -A_1 &0 \\
0 & 0 & -A_3+A^{xy}_4 &0 &0 &A_3 \\
0 & 0 & 0 & 2A^x_5& 0 &0 \\
0 & 0 & 0 &0 & 2A^y_5 & 0 \\
0 & 0 & 0 & 0 & 0 & A^x_5+A^y_5
\end{array}\right].
\label{eq:Rmatrix}
\end{flalign}
In that case, the non-zero two-point correlations that exist are $E_x$ and $E_y$ whose time evolution is given by Eq.~(\ref{time-ev}). In the new set of variables $E_{tot}=E_x+E_y$ and $E_{dif}=E_x-E_y$, the time evolutions reduces to Eq.~(\ref{time ev E}) whose solution is given by Eq.~(\ref{sol E1 one driven}) where the coefficients $K_+, K_-, L_+$ and $L_-$ along with $E^{st}_{tot}$ and $E^{st}_{dif}$ are given by
\begin{widetext}
\begin{align}
&K_+=\frac{1}{\gamma}\Big[ -(\lambda_--\chi_{11})E_{tot}(0)+\chi_{12}E_{dif}(0)-\frac{\lambda_d}{\lambda_+}\big[\big( \chi_{12}-(\lambda_--\chi_{11})  \big)\sigma^2_x - \big( \chi_{12}+(\lambda_--\chi_{11})  \big) \sigma^2_y \big] \Big], \nonumber\\
&K_-= \frac{1}{\gamma}\Big[ (\lambda_+-\chi_{11})E_{tot}(0)-\chi_{12}E_{dif}(0)+ \frac{\lambda_d}{\lambda_-}\big[\big( \chi_{12}-(\lambda_+-\chi_{11})  \big)\sigma^2_x - \big( \chi_{12}+(\lambda_+-\chi_{11})  \big) \sigma^2_y \big] \Big],\nonumber \\
& E^{st}_{tot}=\frac{\lambda_d}{\gamma}\Big[ \frac{\big( \chi_{12}-(\lambda_--\chi_{11})  \big)\sigma^2_x - \big( \chi_{12}+(\lambda_--\chi_{11})  \big) \sigma^2_y}{\lambda_+} - \frac{\big( \chi_{12}-(\lambda_+-\chi_{11})  \big)\sigma^2_x - \big( \chi_{12}+(\lambda_+-\chi_{11})  \big) \sigma^2_y}{\lambda_-} \Big],\nonumber\\
&L_+=\frac{1}{\gamma}\Big[ -\frac{(\lambda_+-\chi_{11})(\lambda_--\chi_{11})}{\chi_{12}}E_{tot}(0) + (\lambda_+ - \chi_{11})E_{dif}(0) -\frac{\lambda_d}{\lambda_+ \chi_{12}}\big[(\lambda_+-\chi_{11})(\lambda_--\chi_{11}) (\sigma^2_x- \sigma^2_y) \big] \Big], \nonumber\\ 
&L_-=\frac{1}{\gamma}\Big[ \frac{(\lambda_+-\chi_{11})(\lambda_--\chi_{11})}{\chi_{12}}E_{tot}(0) - (\lambda_+ - \chi_{11})E_{dif}(0) +\frac{\lambda_d}{\lambda_- \chi_{12}}\big[(\lambda_+-\chi_{11})(\lambda_--\chi_{11}) (\sigma^2_x- \sigma^2_y) \big] \Big], \nonumber\\ 
& E^{st}_{dif}=\frac{\lambda_d}{\chi_{12}\gamma} \Big[ (\lambda_+-\chi_{11})(\lambda_--\chi_{11}) (\sigma^2_x- \sigma^2_y) \big( \frac{1}{\lambda_+}-\frac{1}{\lambda_-} \big) \Big]            , \nonumber\\ 
&\gamma=\lambda_+-\lambda_-.    \label{appendix coeff both components driven}
\end{align}
\end{widetext}

\section{ Hard disc granular gas \label{appendix 2}}

\subsection{ Form of the collision integral}

The time evolution of the velocity distribution function $f(\boldsymbol{v},t)$ for an anisotropically driven hard disc granular is described using the Enskog-Boltzmann equation~\cite{Noije:98}
\begin{align}
\frac{\partial}{\partial t}f(\boldsymbol{{v}},t)=\chi I(f,f) + \Big(\frac{\sigma^2_{x}}{2} \frac{\partial^2}{\partial v^2_x} + \frac{\sigma^2_{y}}{2} \frac{\partial^2}{\partial v^2_y}\Big) f(\boldsymbol{{v}},t), \label{appendix:boltzmann eq}
\end{align}
where $\chi$~\cite{brilliantov2010kinetic} is the pair correlation function, $I(f,f)$ is the collision integral which is given by
\begin{align}
&I(f,f)=\sigma \int d \boldsymbol{{v}}_2 \int d \boldsymbol{{e}} \Theta(-\boldsymbol{{v}}_{12}.\boldsymbol{{e}}) \nonumber \\
& \left|\boldsymbol{{v}}_{12}.\boldsymbol{{e}} \right| \Big[ \frac{1}{r^2} f(\boldsymbol{v}^{''}_1,t) f(\boldsymbol{v}^{''}_2,t) - f(\boldsymbol{{v}}_1,t) f(\boldsymbol{{v}}_2,t)  \Big], \label{collision integral}
\end{align}
where $(\boldsymbol{v}^{''}_1,\boldsymbol{v}^{''}_2)$ are the pre-collision velocities that give $(\boldsymbol{v}_1, \boldsymbol{v}_2)$ upon collision.

\subsection{ Time evolution of $E_{tot}$ and $E_{dif}$}
In this appendix, we write the explicit expressions for the time evolution of $E_{tot}$ and $E_{dif}$ for the hard disc granular gas. The time evolutions of $E_{tot}(t)$ and $E_{dif}(t)$ are written in a compact as given in Eq.~(\ref{time ev}) where the functions $\mathcal{F}(E_{tot},E_{dif})$ and $\mathcal{G}(E_{tot},E_{dif})$ are given by
\begin{widetext}
\begin{align}
&\mathcal{F}(E^{st}_{tot},E^{st}_{dif}) = {\frac{m (\xi^2_{0x} + \xi^2_{0y})}{\nu_0}} + \frac{n \chi \sigma (1+r) \sqrt{E^{st}_{tot} - E^{st}_{dif}} }{15 \nu_0  \sqrt{2 \pi m} E^{st}_{dif}} \times \Bigg[ \Big((3r-7)T^2_{tot} + 4(7r-3) E^{st}_{tot} E^{st}_{dif} + 3(3r-7)T^2_{dif} \Big) \nonumber\\
&{\textit{E}}\Big(\frac{-2E^{st}_{dif}}{E^{st}_{tot} - E^{st}_{dif}}\Big) 
-(3r-7)(E^{st}_{tot} - 3E^{st}_{dif}) \sqrt{T^2_{tot} - T^2_{dif}} \textit{{E}}\Big(\frac{2E^{st}_{dif}}{E^{st}_{tot} + E^{st}_{dif}}\Big) 
- (E^{st}_{tot} + E^{st}_{dif}) \Big((3r-7) E^{st}_{tot} + (7r-3) E^{st}_{dif} \Big)  \nonumber\\
&\textit{{K}}\Big(\frac{-2E^{st}_{dif}}{E^{st}_{tot} - E^{st}_{dif}}\Big)  
 + (3r-7) \sqrt{E^{st}_{tot} + E^{st}_{dif}} (E^{st}_{tot} - E^{st}_{dif})^{3/2}  \textit{{K}}\Big(\frac{2E^{st}_{dif}}{E^{st}_{tot} + E^{st}_{dif}}\Big)  \Bigg], \label{time ev Tt}\\
&\mathcal{G}(E^{st}_{tot},E^{st}_{dif}) ={\frac{m(\xi^2_{0x} - \xi^2_{0y})}{\nu_0}} + \frac{ n \chi \sigma (1+r) (3r-7) \sqrt{E^{st}_{tot} - E^{st}_{dif}} }{15 \nu_0  \sqrt{2 \pi m} E^{st}_{dif}}  
\times \Bigg[(E^{st}_{tot} + 3E^{st}_{dif})(E^{st}_{tot} + E^{st}_{dif}) \textit{{E}}\Big(\frac{-2E^{st}_{dif}}{E^{st}_{tot} - E^{st}_{dif}}\Big)  \nonumber \\
&+(E^{st}_{tot} - 3E^{st}_{dif}) \sqrt{T^2_{tot} - T^2_{dif}} \textit{{E}}\Big(\frac{2E^{st}_{dif}}{E^{st}_{tot} + E^{st}_{dif}}\Big)  
 - (E^{st}_{tot} + E^{st}_{dif})^2   \textit{{K}}\Big(\frac{-2E^{st}_{dif}}{E^{st}_{tot} - E^{st}_{dif}}\Big)  \nonumber \\ 
& - \sqrt{E^{st}_{tot} + E^{st}_{dif}} (E^{st}_{tot} - E^{st}_{dif})^{3/2}  \textit{{K}}\Big(\frac{2E^{st}_{dif}}{E^{st}_{tot} + E^{st}_{dif}}\Big)  \Bigg], \label{time ev Td}
\end{align}
\end{widetext}
where $K(x)$ and $E(x)$ are elliptic integrals of first and second kind respectively. Here, time $t$ is measured in terms of $1/\nu_0$ such that it is dimensionless where $\nu_0$ is the frequency of interparticle collisions given by
\bea
\nu_0=\chi \sigma n \sqrt{\frac{2 (E^{st}_{tot}+E^{st}_{dif})}{\pi m}} E\Big(\frac{2 E^{st}_{dif}}{E^{st}_{tot} + E^{st}_{dif}} \Big).
\eea

However, in order to make the analysis analytically tractable, we linearize the non-linear time evolution equations in (\ref{time ev}) by considering initial states that are close to final stationary state denoted by $E^{st}_{tot}$ and $E^{st}_{dif}$. In that case, the linearized time evolutions are written in a compact form as given in Eq.~(\ref{time ev delta T}) where the elements of the matrix $\boldsymbol{\chi}$ are given by
\begin{widetext}
\bea
\begin{split}
\chi_{11}=&-\frac{n \chi \sigma (1+r) }{30 \nu_0  \sqrt{2 \pi m} E^{st}_{dif}\sqrt{E^{st}_{tot} - E^{st}_{dif}}} \times
\Big[((3 - 47 r) (E^{st}_{dif})^2 + 8 (9r -1) E^{st}_{dif} E^{st}_{tot} + 5( 3 r-7) (E^{st}_{tot})^2)  \textit{{E}}\Big(\frac{-2E^{st}_{dif}}{E^{st}_{tot} - E^{st}_{dif}}\Big)   \\ 
&+\frac{7-3r}{\sqrt{(E^{st}_{tot})^2 - (E^{st}_{dif})^2}}\Big(5 (E^{st}_{dif})^3 + 3 (E^{st}_{dif})^2 E^{st}_{tot} -13 E^{st}_{dif}(E^{st}_{tot})^2 + 5 (E^{st}_{tot})^3 \Big) \textit{{E}}\Big(\frac{2E^{st}_{dif}}{E^{st}_{tot} + E^{st}_{dif}}\Big)    \\
&+ \frac{(7-3r) (E^{st}_{tot} - E^{st}_{dif})^{3/2}}{\sqrt{E^{st}_{tot} + E^{st}_{dif}}} \Big(E^{st}_{dif} (-4\sqrt{E^{st}_{tot} + E^{st}_{dif}} +1) - E^{st}_{tot}(2\sqrt{E^{st}_{tot} + E^{st}_{dif}} +3) \Big)  \textit{{K}}\Big(\frac{2E^{st}_{dif}}{E^{st}_{tot} + E^{st}_{dif}}\Big)   \\
& +\Big((13r-17)(E^{st}_{dif})^2 + 2(1-9r)E^{st}_{dif} E^{st}_{tot} + 5(7-3r)(E^{st}_{tot})^2 \Big)  \textit{{K}}\Big(\frac{-2E^{st}_{dif}}{E^{st}_{tot} - E^{st}_{dif}}\Big)    \\
& +\frac{\pi E^{st}_{dif} }{2 (E^{st}_{tot} - E^{st}_{dif})} \Big(3(3r-7)(E^{st}_{dif})^2 + 4(7r-3) E^{st}_{dif} E^{st}_{tot} + (3r-7) (E^{st}_{tot})^2 \Big)  {}_2F_1\Big(\frac{1}{2},\frac{3}{2};2; \frac{-2E^{st}_{dif}}{E^{st}_{tot} - E^{st}_{dif}} \Big)   \\
& -\frac{\pi (3r-7) E^{st}_{dif} \sqrt{E^{st}_{tot} - E^{st}_{dif}} }{2 (E^{st}_{tot} + E^{st}_{dif})^{3/2}} \Big(3(E^{st}_{dif})^2 - 4 E^{st}_{dif} E^{st}_{tot} +  (E^{st}_{tot})^2 \Big) {}_2F_1\Big(\frac{1}{2},\frac{3}{2};2; \frac{2E^{st}_{dif}}{E^{st}_{tot} + E^{st}_{dif}} \Big)   \\
& - \frac{\pi E^{st}_{dif} (E^{st}_{tot} + E^{st}_{dif}) }{2 (E^{st}_{tot} - E^{st}_{dif})} \Big((7r-3) E^{st}_{dif} + (3r-7) E^{st}_{tot} \Big) {}_2F_1\Big(\frac{3}{2},\frac{3}{2};2; \frac{-2E^{st}_{dif}}{E^{st}_{tot} - E^{st}_{dif}} \Big)    \\
&-\frac{\pi (3r-7) E^{st}_{dif} (E^{st}_{tot} - E^{st}_{dif})^{5/2} }{2 (E^{st}_{tot} + E^{st}_{dif})^{3/2}} {}_2F_1\Big(\frac{3}{2},\frac{3}{2};2; \frac{2E^{st}_{dif}}{E^{st}_{tot} + E^{st}_{dif}} \Big)
\Big], \label{R11}
\end{split}
\eea
\bea
\begin{split}
\chi_{12}=&-\frac{n \chi \sigma (1+r) }{30 \nu_0  \sqrt{2 \pi m} (E^{st}_{dif})^2\sqrt{E^{st}_{tot} - E^{st}_{dif}}}   \\
& \times
\Big[\Big(9(7 - 3 r) (E^{st}_{dif})^3 -10(r+3) (E^{st}_{dif})^2 E^{st}_{tot}-(7-3r)E^{st}_{dif}(E^{st}_{tot})^2 + 2( 7-3 r) (E^{st}_{tot})^3\Big)  \textit{{E}}\Big(\frac{-2E^{st}_{dif}}{E^{st}_{tot} - E^{st}_{dif}}\Big)   \\ 
&-(7-3r)\frac{(E^{st}_{tot} - E^{st}_{dif})^{3/2}}{\sqrt{E^{st}_{tot} + E^{st}_{dif}}}\Big(15 (E^{st}_{dif})^2 + 9 E^{st}_{dif} E^{st}_{tot} - 2 (E^{st}_{tot})^2 \Big) \textit{{E}}\Big(\frac{2E^{st}_{dif}}{E^{st}_{tot} + E^{st}_{dif}}\Big)    \\
&- \frac{(7-3r) \sqrt{E^{st}_{tot} - E^{st}_{dif}}}{\sqrt{E^{st}_{tot} + E^{st}_{dif}}} \Big(5(E^{st}_{dif})^3 - 4 (E^{st}_{dif})^2 E^{st}_{tot}- 3 (E^{st}_{tot})^2 E^{st}_{dif} + 2 (E^{st}_{tot})^3 \Big)  \textit{{K}}\Big(\frac{2E^{st}_{dif}}{E^{st}_{tot} + E^{st}_{dif}}\Big)   \\
& +\Big(-(3-7r)(E^{st}_{dif})^3 + 4(1+r) (E^{st}_{dif})^2 E^{st}_{tot}+ 5(1-5r) (E^{st}_{tot})^2 E^{st}_{dif} + 2(7-3r)(E^{st}_{tot})^3 \Big)  \textit{{K}}\Big(\frac{-2E^{st}_{dif}}{E^{st}_{tot} - E^{st}_{dif}}\Big)    \\
& -\frac{\pi E^{st}_{dif} E^{st}_{tot} }{2 (E^{st}_{tot} - E^{st}_{dif})} \Big(3(7-3r)(E^{st}_{dif})^2 + 4(3-7r) E^{st}_{dif} E^{st}_{tot} + (7-3r) (E^{st}_{tot})^2 \Big)  {}_2F_1\Big(\frac{1}{2},\frac{3}{2};2; \frac{-2E^{st}_{dif}}{E^{st}_{tot} - E^{st}_{dif}} \Big)   \\
& -\frac{\pi (7-3r) E^{st}_{dif} E^{st}_{tot} \sqrt{E^{st}_{tot} - E^{st}_{dif}} }{2 (E^{st}_{tot} + E^{st}_{dif})^{3/2}} \Big(3(E^{st}_{dif})^2 - 4 E^{st}_{dif} E^{st}_{tot} +  (E^{st}_{tot})^2 \Big) {}_2F_1\Big(\frac{1}{2},\frac{3}{2};2; \frac{2E^{st}_{dif}}{E^{st}_{tot} + E^{st}_{dif}} \Big)   \\
& - \frac{\pi E^{st}_{dif} E^{st}_{tot} (E^{st}_{tot} + E^{st}_{dif}) }{2 (E^{st}_{tot} - E^{st}_{dif})} \Big((3-7r) E^{st}_{dif} + (7-3r) E^{st}_{tot} \Big) {}_2F_1\Big(\frac{3}{2},\frac{3}{2};2; \frac{-2E^{st}_{dif}}{E^{st}_{tot} - E^{st}_{dif}} \Big)    \\
&-\frac{\pi (7-3r) E^{st}_{dif} E^{st}_{tot} (E^{st}_{tot} - E^{st}_{dif})^{5/2} }{2 (E^{st}_{tot} + E^{st}_{dif})^{3/2}} {}_2F_1\Big(\frac{3}{2},\frac{3}{2};2; \frac{2E^{st}_{dif}}{E^{st}_{tot} + E^{st}_{dif}} \Big)
\Big], \label{R12}  
\end{split}
\eea
\bea
\begin{split}
\chi_{21}=&\frac{n \chi \sigma (1+r) (7-3r)}{30 \nu_0  \sqrt{2 \pi m} E^{st}_{dif}\sqrt{E^{st}_{tot} - E^{st}_{dif}}} \times 
\Big[\Big(-5 (E^{st}_{dif})^2 + 8  E^{st}_{dif} E^{st}_{tot} + 5 (E^{st}_{tot})^2 \Big)  \textit{{E}}\Big(\frac{-2E^{st}_{dif}}{E^{st}_{tot} - E^{st}_{dif}}\Big)   \\ 
&+\Big(\frac{(E^{st}_{tot} + E^{st}_{dif})^2}{2}+ (3E^{st}_{tot} - E^{st}_{dif})(E^{st}_{tot} -3 E^{st}_{dif}) \frac{\sqrt{E^{st}_{tot} - E^{st}_{dif}}}{\sqrt{E^{st}_{tot} + E^{st}_{dif}}}\Big) \textit{{E}}\Big(\frac{2E^{st}_{dif}}{E^{st}_{tot} + E^{st}_{dif}}\Big)    \\
&-\Big(\frac{4}{\sqrt{(E^{st}_{tot})^2 - (E^{st}_{dif})}} \big(4 (E^{st}_{dif})^2 -  E^{st}_{dif} E^{st}_{tot} - (E^{st}_{tot})^2 \big) \Big) \textit{{E}}\Big(\frac{2E^{st}_{dif}}{E^{st}_{tot} + E^{st}_{dif}}\Big)    \\
&- \frac{ (E^{st}_{tot} - E^{st}_{dif})^{3/2}}{\sqrt{E^{st}_{tot} + E^{st}_{dif}}} \Big(3 E^{st}_{dif} + 5 E^{st}_{tot} \Big)  \textit{{K}}\Big(\frac{2E^{st}_{dif}}{E^{st}_{tot} + E^{st}_{dif}}\Big) -\Big((5E^{st}_{tot}-3 E^{st}_{dif}) (E^{st}_{tot}+ E^{st}_{dif})\Big)  \textit{{K}}\Big(\frac{-2E^{st}_{dif}}{E^{st}_{tot} - E^{st}_{dif}}\Big)    \\
& -\frac{\pi E^{st}_{dif}  (E^{st}_{tot} +3 E^{st}_{dif})  (E^{st}_{tot} + E^{st}_{dif})} {2 (E^{st}_{tot} - E^{st}_{dif})}   {}_2F_1\Big(\frac{1}{2},\frac{3}{2};2; \frac{-2E^{st}_{dif}}{E^{st}_{tot} - E^{st}_{dif}} \Big)   -\frac{\pi E^{st}_{dif} (E^{st}_{tot} + E^{st}_{dif})^2} {2 (E^{st}_{tot} - E^{st}_{dif})} {}_2F_1\Big(\frac{3}{2},\frac{3}{2};2; \frac{-2E^{st}_{dif}}{E^{st}_{tot} - E^{st}_{dif}} \Big)   \\
& +\frac{\pi E^{st}_{dif}  (E^{st}_{tot} -3 E^{st}_{dif})(E^{st}_{tot} - E^{st}_{dif})^{3/2} }{2 (E^{st}_{tot} + E^{st}_{dif})^{3/2}}  {}_2F_1\Big(\frac{1}{2},\frac{3}{2};2; \frac{2E^{st}_{dif}}{E^{st}_{tot} + E^{st}_{dif}} \Big) +\frac{\pi E^{st}_{dif} (E^{st}_{tot} - E^{st}_{dif})^{5/2}} {2 (E^{st}_{tot} + E^{st}_{dif})^{3/2}} {}_2F_1\Big(\frac{3}{2},\frac{3}{2};2; \frac{2E^{st}_{dif}}{E^{st}_{tot} + E^{st}_{dif}} \Big)
\Big], \label{R21}  
\end{split}
\eea
and 
\bea
\begin{split}
\chi_{22}=&\frac{n \chi \sigma (1+r)(7-3r) }{30 \nu_0  \sqrt{2 \pi m} (E^{st}_{dif})^2\sqrt{E^{st}_{tot} - E^{st}_{dif}}} \times
\Big[\Big(-9 (E^{st}_{dif})^3 + 2 (E^{st}_{dif})^2 E^{st}_{tot}+E^{st}_{dif}(E^{st}_{tot})^2 - 2 (E^{st}_{tot})^3\Big)  \textit{{E}}\Big(\frac{-2E^{st}_{dif}}{E^{st}_{tot} - E^{st}_{dif}}\Big)   \\ 
&+\frac{\sqrt{E^{st}_{tot} - E^{st}_{dif}}}{\sqrt{E^{st}_{tot} + E^{st}_{dif}}}\Big(9 (E^{st}_{dif})^3 + 2 (E^{st}_{dif})^2 E^{st}_{tot}-E^{st}_{dif}(E^{st}_{tot})^2 - 2 (E^{st}_{tot})^3 \Big) \textit{{E}}\Big(\frac{2E^{st}_{dif}}{E^{st}_{tot} + E^{st}_{dif}}\Big)    \\
&+ \frac{ (E^{st}_{tot} - E^{st}_{dif})^{3/2}}{\sqrt{E^{st}_{tot} + E^{st}_{dif}}} \Big(3 (E^{st}_{dif})^2 + 3 E^{st}_{dif} E^{st}_{tot}  + 2 (E^{st}_{tot})^2 \Big)  \textit{{K}}\Big(\frac{2E^{st}_{dif}}{E^{st}_{tot} + E^{st}_{dif}}\Big)   \\
& +\Big(3 (E^{st}_{dif})^2 - 3 E^{st}_{dif} E^{st}_{tot}  + 2 (E^{st}_{tot})^2 \Big)  \textit{{K}}\Big(\frac{-2E^{st}_{dif}}{E^{st}_{tot} - E^{st}_{dif}}\Big)    \\
& +\frac{\pi E^{st}_{dif} E^{st}_{tot} (E^{st}_{tot} + E^{st}_{dif}) (E^{st}_{tot} + 3E^{st}_{dif}) }{2 (E^{st}_{tot} - E^{st}_{dif})}   {}_2F_1\Big(\frac{1}{2},\frac{3}{2};2; \frac{-2E^{st}_{dif}}{E^{st}_{tot} - E^{st}_{dif}} \Big)   \\
& -\frac{\pi E^{st}_{dif} E^{st}_{tot} (E^{st}_{tot} - 3 E^{st}_{dif}) (E^{st}_{tot} - E^{st}_{dif})^{3/2}} {2 (E^{st}_{tot} + E^{st}_{dif})^{3/2}} {}_2F_1\Big(\frac{1}{2},\frac{3}{2};2; \frac{2E^{st}_{dif}}{E^{st}_{tot} + E^{st}_{dif}} \Big)   \\
& +\frac{\pi E^{st}_{dif} E^{st}_{tot} (E^{st}_{tot} + E^{st}_{dif})^2 }{2 (E^{st}_{tot} - E^{st}_{dif})} {}_2F_1\Big(\frac{3}{2},\frac{3}{2};2; \frac{-2E^{st}_{dif}}{E^{st}_{tot} - E^{st}_{dif}} \Big)    \\
&-\frac{\pi E^{st}_{dif} E^{st}_{tot} (E^{st}_{tot} - E^{st}_{dif})^{5/2}} {2 (E^{st}_{tot} + E^{st}_{dif})^{3/2}} {}_2F_1\Big(\frac{3}{2},\frac{3}{2};2; \frac{2E^{st}_{dif}}{E^{st}_{tot} + E^{st}_{dif}} \Big)
\Big], \label{R22}
\end{split}
\eea
\end{widetext}
where ${}_2F_1(a,b;c;z)'s$ are the hypergeometric function.

%

\end{document}